\documentclass[12pt]{article}
\usepackage{geometry}
\usepackage{authblk}
\usepackage{url}
\usepackage{graphicx}%
\usepackage{multirow}%
\usepackage{amsmath,amssymb,amsfonts,bm}%
\usepackage{amsthm}%
\usepackage{mathrsfs}%
\usepackage[title]{appendix}%
\usepackage{xcolor}%
\usepackage{textcomp}%
\usepackage{manyfoot}%
\usepackage{booktabs}%
\usepackage{algorithm}%
\usepackage{algorithmicx}%
\usepackage{algpseudocode}%
\usepackage{listings}%
\usepackage{tabularx}%
\usepackage{mathtools}

\theoremstyle{plain}
\newtheorem{lemma}{Lemma}
\newtheorem{theorem}{Theorem}

\newcolumntype{A}{>{\raggedright\arraybackslash}p{2.6cm}}
\newcolumntype{B}{>{\centering\arraybackslash}p{1.8cm}}
\newcolumntype{C}{>{\centering\arraybackslash}p{1cm}}
\newcolumntype{D}{>{\centering\arraybackslash}p{1.2cm}}
\newcolumntype{E}{>{\centering\arraybackslash}p{1.8cm}}
\newcolumntype{F}{>{\centering\arraybackslash}p{1cm}}

\title{Sampling nodes and hyperedges via random walks on large hypergraphs}

\author[1, *]{Kazuki Nakajima}
\author[1]{Masanao Kodakari}
\author[1]{Masaki Aida}

\affil[1]{Graduate School of Systems Design, Tokyo Metropolitan University, 6-6 Asahigaoka, Hino-shi, 191-0065, Tokyo, Japan}
\affil[*]{Corresponding author: nakajima@tmu.ac.jp}

\begin{document}
\date{}
\maketitle

\begin{abstract}
Hypergraphs provide a fundamental framework for representing complex systems involving interactions among three or more entities. 
As empirical hypergraphs grow in size, characterizing their structural properties becomes increasingly challenging due to computational complexity and, in some cases, restricted access to complete data, requiring efficient sampling methods. 
Random walks offer a practical approach to hypergraph sampling, as they rely solely on local neighborhood information from nodes and hyperedges. 
In this study, we investigate methods for simultaneously sampling nodes and hyperedges via random walks on large hypergraphs. 
First, we compare three existing random walks in the context of hypergraph sampling and identify an advantage of the so-called higher-order random walk. 
Second, by extending an established technique for graphs to the case of hypergraphs, we present a non-backtracking variant of the higher-order random walk. 
We derive theoretical results on estimators based on the non-backtracking higher-order random walk and validate them through numerical simulations on large empirical hypergraphs. 
Third, we apply the non-backtracking higher-order random walk to a large hypergraph of co-authorships indexed in the OpenAlex database, where full access to the data is not readily available.
Despite the relatively small sample size, our estimates largely align with previous findings on author productivity, team size, and the prevalence of open-access publications.
Our findings contribute to the development of analysis methods for large hypergraphs, offering insights into sampling strategies and estimation techniques applicable to real-world complex systems.
\end{abstract}

{\flushleft{{\bf Keywords:} Hypergraphs, higher-order networks, random walk, sampling, estimation, hypergraph sampling, coauthorship hypergraphs}}

\section{Introduction} \label{section:1}

Graphs are a widely accepted representation of real-world complex systems, consisting of a set of entities (i.e., nodes) and a set of dyadic interactions (i.e., edges) between them \cite{newman2018}. 
A number of studies on graph data analysis have contributed to the understanding of the structure and dynamics of complex systems. 
While large graphs such as the World Wide Web and online social networks are now commonplace, managing and analyzing them often poses significant technical challenges. 
It can be computationally expensive to analyze the structure of a large graph in its entirety, such as shortest-path related properties \cite{brandes2008}.
In addition, access to its full snapshot may be limited for reasons unrelated to computational resources. 
For example, online social networks may not be fully visible to the public or may only be accessible through limited interfaces \cite{mislove2007}. 
Measurements or surveys required to observe the underlying social networks of individuals may be costly \cite{salganik2004}. 
A practical and fundamental approach to these cases is to sample a subset of nodes and/or edges in a large graph and estimate its structural properties \cite{leskovec2006, ahmed2013}.

Random walks have various applications in graph analysis \cite{lovasz1993, masuda2017}.
In a simple random walk on a graph, one starts with a given node (e.g., an individual or user) and repeatedly moves to a randomly selected neighboring node.
Random walks are suitable for sampling a large graph to which access is restricted because they can be performed relying on the neighborhood information of each node.
Indeed, they have been used to sample hard-to-reach populations in social surveys (e.g., injection drug users) and estimate their proportion \cite{salganik2004}, to sample Facebook users and estimate properties of the Facebook graph \cite{gjoka2011}, and to sample X (formerly Twitter) users and estimate the proportion of `bot' accounts \cite{fukuda2022}.
Another important advantage of random walks is the theoretical guarantees provided by the Markov property of the sampled node sequence. 
In particular, one obtains unbiased estimators of various graph properties by properly re-weighting each sampled node. 
Based on this, previous studies have developed estimators of graph properties based on random walks (e.g., \cite{gjoka2011, ribeiro2010, hardiman2013, wang2014, chen2016}).

Real-world complex systems often involve higher-order interactions among entities (i.e., interactions among three or more entities).
Examples include group conversations among students in schools \cite{stehl2011, mastrandrea2015}, coauthorships among researchers \cite{patania2017}, threads on question-and-answer websites in which users participate \cite{benson2018}, and many others \cite{battiston2020}.
Hypergraphs, which consist of a set of nodes and a set of hyperedges among two or more nodes, are an alternative representation of these systems.
Owing to recent advances in theories, measurements, computational methods, and dynamical models for hypergraphs, higher-order interactions have been found to play a significant role in describing the structures, phenomena, and dynamics of complex systems \cite{battiston2020, battiston2021, majhi2022, boccaletti2023}.
As the scale of empirical hypergraphs grows, the computational complexity of characterizing their structures becomes a practical issue \cite{ruggeri2023, lotito2024}.
Additionally, as in the case of large publication databases \cite{sinha2015, priem2022}, access to complete data on underlying hypergraphs can be limited through public interfaces.
Thus motivated, previous studies have recently investigated the hypergraph sampling problem, focusing on how to effectively sample nodes and hyperedges in large hypergraphs \cite{choe2022, choe2024, lee2025}.

In hypergraphs, random walks can be defined in various ways, depending on the choice of state space and the transition rules between states. 
Previous studies considered the set of nodes as the state space and defined various transition rules between two nodes that may share multiple hyperedges of different cardinalities \cite{zhou2006, chaudhuri2019, hayashi2020, aksoy2020, battiston2020, carletti2020, eriksson2021, carletti2021, banerjee2021, nagasato2023, traversa2024}. 
Other work instead used a state space consisting of node-hyperedge pairs and explored how to simultaneously sample nodes and hyperedges via random walks \cite{zhang2023, luo2024}. 
However, the understanding of these diverse random walks from a hypergraph sampling perspective remains limited. 
In practice, we need to consider the constraints on the number of queries required to obtain neighborhood information from nodes and hyperedges, similar to the case of graphs \cite{gjoka2011, hardiman2013}. 
Additionally, techniques used to improve the accuracy of estimators based on random walks in graphs (e.g., \cite{lee2012, li2015}) could potentially be extended to hypergraphs. 
Moreover, there are few case studies examining the application of random walks to hypergraphs when the complete snapshot is unavailable.

In this study, we focus on sampling nodes and hyperedges via random walks and estimating structural properties of large hypergraphs. 
We make three main contributions. 
First, using a general framework of random walks on a hypergraph \cite{zhou2006, hayashi2020}, we compare three existing random walks in the context of hypergraph sampling. 
We find that the higher-order random walk \cite{banerjee2021, luo2024, traversa2024} is more efficient in terms of the number of queries to nodes and hyperedges than other random walks. 
Second, we extend the higher-order random walk by adopting the non-backtracking technique \cite{lee2012, li2015} from graphs to hypergraphs. 
We discuss the convergence of estimators of node and hyperedge properties based on the non-backtracking higher-order random walk. 
We conduct numerical simulations on empirical hypergraphs to validate these estimators.
Third, we showcase the application of the non-backtracking higher-order random walk to a hypergraph composed of hundreds of millions of scientific coauthorships, where the full snapshot is not readily available. 
This work is a significantly extended version of our previous work \cite{kodakari2025, kodakari20252} because this work includes these new contributions.
Our code is available at \url{https://github.com/kazuibasou/hrw_sampling}.

\section{Methods} \label{section:2}

\subsection{Hypergraph} \label{section:2.1}

We represent a hypergraph $H$ by a set of nodes and a set of hyperedges among nodes.
Let $V=\{v_1, \,\dots, , v_n\}$ be the set of nodes and $E=\{e_1,\,\dots,\,e_m\}$ be the set of hyperedges, where $n$ and $m$ are the numbers of nodes and hyperedges, respectively.
We assume that any hyperedge $e_\alpha \in E$ is a subset of $V$ (i.e., $e_\alpha \subseteq V$) and its cardinality is two or larger (i.e., $|e_\alpha| \geq 2$).
We define an $n \times m$ incidence matrix, $\bm{B} = (b_{i,\alpha})_{1 \leq i \leq n, 1 \leq \alpha \leq m}$, where $b_{i,\alpha} = 1$ if node $v_i$ belongs to hyperedge $e_\alpha$ and $b_{i,\alpha} = 0$ otherwise.
We denote by $\Gamma_e(i)$ the set of hyperedges to which node $v_i \in V$ belongs and by $\Gamma_v(\alpha)$ the set of nodes belonging to hyperedge $e_{\alpha}$.
We define the degree of node $v_i$ as $d_i = \sum_{\alpha=1}^m b_{i, \alpha} = |\Gamma_e(i)|$, which is the number of hyperedges to which node $v_i$ belongs.
We define the size of the hyperedge of $e_{\alpha}$ as $s_{\alpha} = \sum_{i=1}^n b_{i, \alpha} = |\Gamma_v(\alpha)|$, which is the number of nodes that belong to hyperedge $e_{\alpha}$.
We also assume that $H$ is fully connected (i.e., forms a single connected component), where two nodes $u$ and $v$ are connected if a walker can move from $u$ to $v$ by repeatedly traversing hyperedges \cite{zhou2006}.

\subsection{Sampling of nodes and hyperedges} \label{section:2.2}

\subsubsection{Problem definition}

Given a positive integer $r$, we sample $r$ node indices and $r$ hyperedge indices with replacement via a random walk on the hypergraph $H$. 
By extending the standard assumptions for graphs \cite{gjoka2010,nakajima2023}, we assume the following about $H$: (i) when querying a node $v_i \in V$, we obtain the set of hyperedges to which $v_i$ belongs, i.e., $\Gamma_e(i)$; (ii) when querying a hyperedge $e_{\alpha} \in E$, we obtain the set of nodes that belong to $e_{\alpha}$, i.e., $\Gamma_v(\alpha)$; (iii) $H$ is static; and (iv) a seed node is chosen properly to perform a random walk on $H$. 
In practical scenarios, the execution of a random walk is restricted by the number of queries to nodes and hyperedges \cite{hardiman2013, gjoka2011}. 
In addition, the quality of a sample of \(H\) can be evaluated by comparing certain properties of \(H\) with their estimated values derived from the sample \cite{choe2022, choe2024, lee2025}. 
Therefore, we seek a random walk that minimizes the number of queries to nodes and hyperedges while producing estimates of certain node and hyperedge properties of \(H\) that are as accurate as possible.

\subsubsection{Existing random walks} \label{section:2.2.2}

We apply a general framework for random walks on hypergraphs \cite{zhou2006, hayashi2020} to our problem. 
Given a seed node, we generate a random walk on a hypergraph by repeatedly sampling a connected hyperedge and subsequently sampling a different node contained within this hyperedge. 
We describe the algorithmic procedure below, where \( X_k \) and \( Y_k \) denote the indices of the \( k \)-th sampled node and hyperedge, respectively.

\begin{enumerate}
\item Create an empty list, $L$. Choose a seed node, $v_{X_1} \in V$. Initialize $k$ with one.
\item Query the node $v_{X_k}$ to get the set $\Gamma_e(X_{k})$. Append $X_k$ to $L$. Sample a hyperedge $e_{Y_{k}} \in \Gamma_e(X_{k})$ according to the probability $S_{X_k, Y_k} \coloneq P(Y_k=\alpha\ |\ X_k=i)$, which is specifically defined for a given random walk.
\item Query the hyperedge $e_{Y_k}$ to get the set $\Gamma_v(Y_{k})$. Append $Y_k$ to $L$. Sample a node $v_{X_{k+1}} \in \Gamma_v(Y_{k})\ \backslash\ \{v_{X_k}\}$ uniformly at random. Increase $k$ by one. If $k < r$, return to action 2; otherwise, stop.
\end{enumerate}

This framework describes three existing random walks: the projected random walk (P-RW) \cite{zhou2006, battiston2020, zhang2023, traversa2024}, the random walk proposed by Carletti et al. (C-RW) \cite{carletti2020}, and the higher-order random walk (HO-RW) \cite{banerjee2021, luo2024, traversa2024}.
We describe the differences among these random walks in terms of (i) the transition probability between nodes, $T_{i,j} \coloneq P(X_{k+1}=j\ |\ X_k = i)$, which has been derived in previous studies, and (ii) the number of queries to nodes and hyperedges.


We first focus on the P-RW, which is a random walk on the projected graph of $H$ (i.e., the weighted graph in which two nodes are connected by an edge whose weight is the number of hyperedges they share in $H$) \cite{zhou2006, battiston2020, zhang2023, traversa2024}.
To define this random walk, we set $S_{i, \alpha} = (s_{\alpha} - 1) \, b_{i,\alpha} / \left[\sum_{\beta=1}^m (s_{\beta} - 1) \, b_{i,\beta}\right]$ for any $i=1, \ldots, n$ and any $\alpha = 1, \ldots, m$.
Indeed, the probability $T_{i,j}$ is given by
\begin{align}
T_{i,j} &= \sum_{\alpha=1}^m P(X_{k+1} = j \ |\ Y_k=\alpha \land X_k=i) \, P(Y_k=\alpha\ |\ X_k = i) \notag \\
&= \sum_{\alpha=1}^m \frac{b_{j,\alpha}}{s_{\alpha} - 1} \, S_{i, \alpha} \notag \\
&= \frac{\sum_{\alpha=1}^m b_{i, \alpha} \, b_{j, \alpha}} {\sum_{\beta=1}^m (s_{\beta} - 1) \, b_{i, \beta}}
\label{eq:1}
\end{align}
for any $1 \leq i,\, j \leq n$ such that $i \neq j$, and $T_{i,i} = 0$ for any $i = 1, \ldots, n$, which is consistent with Refs.~\cite{zhou2006, battiston2020, zhang2023, traversa2024}.
Equation~\eqref{eq:1} implies that a walker being at $v_{X_k} = v_i$ is more likely to visit $v_{X_{k+1}} = v_j$ if they share a larger number of hyperedges.
At each step, P-RW generates a query to the node $v_{X_k}$ and a query to each hyperedge in $\Gamma_e(X_{k})$ to compute the probability $S_{X_k, \alpha}$ for each $e_{\alpha} \in E$.


To define the C-RW \cite{carletti2020}, for any $i=1, \ldots, n$ and any $\alpha = 1, \ldots, m$ we set $S_{i, \alpha} = (s_{\alpha} - 1)^2 \, b_{i,\alpha} / \left[\sum_{\beta=1}^m (s_{\beta} - 1)^2 \, b_{i,\beta}\right]$.
Indeed, it holds true that
\begin{align}
T_{i,j} = \frac{\sum_{\alpha=1}^m (s_{\alpha} - 1) \, b_{i, \alpha} \, b_{j, \alpha}} {\sum_{\beta=1}^m (s_{\beta} - 1)^2 \, b_{i, \beta}}
\label{eq:2}
\end{align}
for any $1 \leq i,\, j \leq n$ such that $i \neq j$, and $T_{i,i} = 0$ for any $i = 1, \ldots, n$, which is consistent with Ref.~\cite{carletti2020}.
Equation~\eqref{eq:2} implies that a walker being at $v_{X_k} = v_i$ is more likely to visit $v_{X_{k+1}} = v_j$ if they share hyperedges of larger sizes.
As with P-RW, at each step, C-RW generates a query to the node $v_{X_k}$ and a query to each hyperedge in $\Gamma_e(X_{k})$.


To define the HO-RW \cite{banerjee2021, luo2024, traversa2024}, we set $S_{i, \alpha} = b_{i, \alpha} / d_i$ for any $i=1, \ldots, n$ and any $\alpha = 1, \ldots, m$.
Then, it holds true that
\begin{align}
T_{i,j} &= \frac{1}{d_i} \sum_{\alpha=1}^m \frac{b_{i,\alpha} \, b_{j,\alpha}}{s_{\alpha} - 1}
\label{eq:3}
\end{align}
for any $1 \leq i, j \leq n$ such that $i \neq j$, and $T_{i,i} = 0$ for any $i = 1, \ldots, n$, which is consistent with Refs.~\cite{banerjee2021, luo2024, traversa2024}.
Equation~\eqref{eq:3} implies that a walker being at $v_{X_k} = v_i$ is more likely to visit $v_{X_{k+1}} = v_j$ if they share hyperedges of smaller sizes.
In contrast to P-RW and C-RW, it is sufficient to query the node $v_{X_k}$ and the hyperedge $e_{Y_k}$ at each step of HO-RW because HO-RW samples a hyperedge from $\Gamma_e(X_{k})$ uniformly at random.
Therefore, HO-RW generates a much smaller number of queries compared to P-RW and C-RW.

\subsubsection{Non-backtracking higher-order random walk}

\begin{figure}[t]
    \centering
    \includegraphics[width=0.6\linewidth]{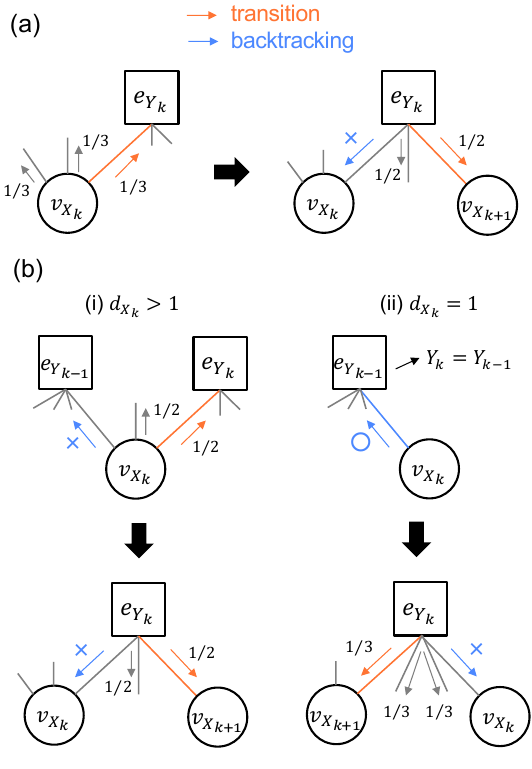}\\
    \caption{A schematic comparison between (a) HO-RW and (b) NB-HO-RW.}
\label{fig:1}
\end{figure}

In general, `backtracking' in a random walk on a conventional graph is defined as the case where \( X_{k+1} = X_{k-1} \) holds true for some \( 2 \leq k \leq r-1 \) \cite{lee2012, li2015}.  
We extend this definition from graphs to hypergraphs because we sample nodes and hyperedges simultaneously via a random walk on \( H \).
Specifically, we define backtracking in a random walk on \( H \) as the case where \( Y_{k} = Y_{k-1} \) for some \( 2 \leq k \leq r \).  
This definition recovers the conventional definition for simple graphs (i.e., graphs containing neither multiple edges nor self-loops, as assumed in Refs.~\cite{hardiman2013, chen2016, nakajima2023}), because it holds that \( X_{k+1} = X_{k-1} \) if and only if \( Y_{k} = Y_{k-1} \) for some \( 2 \leq k \leq r-1 \).
Such backtracking can reduce the accuracy of estimators based on random walks \cite{lee2012, li2015}.
We extend HO-RW to a random walk that avoids backtracking except in unavoidable cases, which we refer to as the non-backtracking higher-order random walk (NB-HO-RW).
By modifying the above framework, we design NB-HO-RW as follows (see also Fig.~\ref{fig:1}).

\begin{enumerate}
\item Create an empty list, $L$. Choose a seed node, $v_{X_1}$. 
\item Query the node $v_{X_1}$ to get the set $\Gamma_e(X_{1})$. Append $X_1$ to $L$. Sample a hyperedge $e_{Y_{1}} \in \Gamma_e(X_{1})$ uniformly at random. 
\item Query the hyperedge $e_{Y_1}$ to get the set $\Gamma_v(Y_{1})$. Append $Y_1$ to $L$. Sample a node $v_{X_{2}} \in \Gamma_v(Y_{1})\ \backslash\ \{v_{X_1}\}$ uniformly at random. Initialize $k$ with two.
\item Query the node $v_{X_k}$ to get the set $\Gamma_e(X_{k})$. Append $X_k$ to $L$. If $d_{X_k} = 1$, set $Y_{k} = Y_{k-1}$. Otherwise, sample a hyperedge $e_{Y_{k}} \in \Gamma_e(X_{k})\ \backslash\ \{e_{Y_{k-1}}\}$ uniformly at random.
\item Query the hyperedge $e_{Y_k}$ to get the set $\Gamma_v(Y_{k})$. Append $Y_k$ to $L$. Sample a node $v_{X_{k+1}} \in \Gamma_v(Y_{k})\ \backslash\ \{v_{X_k}\}$ uniformly at random. Increase $k$ by one. If $k < r$, return to action 4; otherwise, stop.
\end{enumerate}

NB-HO-RW requires no additional queries to nodes or hyperedges beyond those required by HO-RW because it is sufficient to query the node $v_{X_k}$ and the hyperedge $e_{Y_k}$ for each $k = 1, \ldots, r$.
Zhang et al.~extended P-RW to random walks that avoid backtracking \cite{zhang2023}. 
However, P-RW-based random walks are much more expensive than HO-RW in terms of the number of queries to hyperedges, as we described.
Thus, we will focus on estimators of node and hyperedge properties based on HO-RW and NB-HO-RW.
Luo et al.~developed estimators based on HO-RW and presented their theoretical basis \cite{luo2024}.
In the next section, we design estimators based on NB-HO-RW and discuss the convergence of these estimators.

\subsection{Estimation of node and hyperedge properties}

Suppose NB-HO-RW has generated a sample sequence, $L = \{X_1, Y_1, \ldots, X_r, Y_r\}$. 
First, we show the strong law of large numbers associated with this sample sequence.
Then, we describe estimators for node and hyperedge properties derived from the sample sequence and discuss the convergence of these estimators.

\subsubsection{Strong Law of Large Numbers}

The following strong law of large numbers for a Markov chain serves as a theoretical basis for Markov chain Monte Carlo samplers \cite{ribeiro2010,lee2012,li2015,chen2016}.
Consider a Markov chain on a finite state space \(\Omega\), and let \(\bm{P} = (P_{i,j})_{i \in \Omega, j \in \Omega}\) be its transition probability matrix.
The stationary distribution \(\bm{\pi} \coloneq (\pi_i)_{i \in \Omega}\) is defined as a row vector with nonnegative entries that sum to 1, satisfying \(\bm{\pi}^\top = \bm{\pi}^\top \bm{P}\).
If the chain is irreducible and aperiodic, then this stationary distribution exists uniquely (see Ref.~\cite{levin2017} for formal definitions and the proof).
\begin{theorem}
\cite{roberts2004, jones2004}
Suppose that $\{Z_k\}_{k=1}^r$ is a finite and irreducible Markov chain with the state space $\Omega$ and the stationary distribution $\bm{\pi}$. 
Then, for any initial distribution, $P(Z_1 = i')$ for any $i' \in \Omega$, as $r \to \infty$, 
\begin{align*}
\frac{1}{r} \sum_{k=1}^r f(Z_k) \to \sum_{i \in \Omega} f(i) \, \pi_i\ \text{almost surely}
\end{align*}
for any function $f$ such that $\sum_{i \in \Omega} f(i) \pi_i < \infty$.
\label{theorem:1}
\end{theorem}

The sequence $L$ itself is not a Markov chain because $X_{k+1}$ depends on both $X_k$ and $Y_k$.
Instead, we reconstruct the sequence $L' \coloneq \{(X_1, Y_1), (X_2, Y_2), \ldots, (X_r, Y_r)\}$ from $L$.
If and only if $H$ is fully connected, which we assumed in Section \ref{section:2.1}, the sequence $L'$ is a finite, irreducible, and aperiodic Markov chain on the finite state space $\mathcal{E} \coloneq \{(i, \alpha)\ |\ 1 \leq i \leq n\ \land \ 1 \leq \alpha \leq m\ \land\ b_{i, \alpha} = 1\}$ and the transition probability matrix $\bm{U} \coloneq (U_{(i, \alpha), (j, \beta)})_{(i, \alpha) \in \mathcal{E}, (j, \beta) \in \mathcal{E}}$, where we define $U_{(i, \alpha), (j, \beta)} \coloneq P(X_{k+1} = j \land Y_{k+1} = \beta\ |\ X_k = i \land Y_k = \alpha)$.
We have the following lemmas (see Appendices~\ref{appendix:a} and \ref{appendix:b} for the proofs of Lemmas~\ref{lemma:1} and \ref{lemma:2}, respectively).

\begin{lemma}
For any $(i, \alpha) \in \mathcal{E}$ and any $(j, \beta) \in \mathcal{E}$, 
\begin{align*}
U_{(i, \alpha), (j, \beta)} =
\displaystyle \begin{dcases}
\frac{b_{j, \alpha} \, \bm{1}_{\{i \neq j \land \alpha = \beta\}}}{s_{\alpha} - 1}    &   \text{if $d_j = 1$}, \\
\frac{b_{j, \alpha} \, \bm{1}_{\{i \neq j \land \alpha \neq \beta\}}}{(d_j-1)(s_{\alpha} - 1)} &   \text{if $d_j > 1$}.
\end{dcases}
\end{align*}
\label{lemma:1}
\end{lemma}

\begin{lemma}
The stationary distribution of the chain is given by  $\bm{\pi}_{\mathcal{E}} \coloneq (1 / D)_{(i, \alpha) \in \mathcal{E}}$, where $D \coloneq |\mathcal{E}| = \sum_{i=1}^n d_i = \sum_{\alpha=1}^m s_{\alpha}$.
\label{lemma:2}
\end{lemma}

Theorem \ref{theorem:1} and Lemmas \ref{lemma:1} and \ref{lemma:2} imply the following theorem, which serves as the theoretical basis for the convergence of estimators based on NB-HO-RW.
\begin{theorem}
The sequence $L' = \{(X_k, Y_k)\}_{k=1}^{r}$ is a finite, irreducible, and aperiodic Markov chain with the finite state space $\mathcal{E}$ and the stationary distribution $\bm{\pi}_{\mathcal{E}}$ if and only if $H$ is fully connected.
Then, for any initial distribution, $P((X_1, Y_1) = (i', \alpha'))$ for any $(i', \alpha') \in \mathcal{E}$, as $r \to \infty$, 
\begin{align*}
\frac{1}{r} \sum_{k=1}^r g((X_k, Y_k)) \to \sum_{(i, \alpha) \in \mathcal{E}} g((i, \alpha)) \, \pi_{\mathcal{E}, (i, \alpha)}\ \text{almost surely}
\end{align*}
for any function $g: \mathcal{E} \to \mathbb{R}$ such that $\sum_{(i, \alpha) \in \mathcal{E}} g((i, \alpha)) \pi_{\mathcal{E}, (i, \alpha)} < \infty$.
\label{theorem:2}
\end{theorem}

\subsubsection{Node properties}

We estimate the node's property of interest in $H$ defined as $\mu_{v} \coloneqq \sum_{i=1}^n f_v(i) / n$, where $f_v: \{1, \ldots, n\} \to \mathbb{R}$ is a function that returns the feature of interest of node $v_i \in V$.
By appropriately choosing $f_v$, one can specify the desired node feature.
For instance, if $f_v(i) = d_i$, then $\mu_{v}$ is the average degree of a node in $H$ \cite{gjoka2011, lee2012}. 
If \(f_v(i)\) returns 1 when node \(v_i\) has degree \(d\) (and 0 otherwise), then \(\mu_{v}\) equals the probability that a node has degree \(d\); summing these probabilities over all possible \(d\) values yields an estimator for the probability distribution of the node degree \cite{gjoka2011, lee2012}.
If $f_v(i)$ returns 1 if node $v_i$ belongs to a specific certain subpopulation and returns 0 otherwise, then $\mu_{v}$ is the proportion of that subpopulation in the entire network \cite{salganik2004, gjoka2011, fukuda2022}.
We define an estimator of $\mu_{v}$ as 
\begin{align}
\hat{\mu}_{v, r} \coloneqq \Phi_{v, r} / \Psi_{v, r},
\label{eq:4}
\end{align}
where
\begin{align*}
\Phi_{v, r} &= \frac{1}{r} \sum_{k=1}^r \frac{f_v(X_k)}{d_{X_k}}, \\
\Psi_{v, r} &= \frac{1}{r} \sum_{k=1}^r \frac{1}{d_{X_k}}.
\end{align*}
This estimator can be extended to the property of interest for the nodes in a subset $V' \subset V$ (e.g., the average degree of a node in $V'$).
Let us estimate $\mu_{v, V'} \coloneqq \sum_{i=1, v_i \in V'}^n f_v(i) / |V'|$.
We define an estimator of $\mu_{v, V'}$
\begin{align}
\hat{\mu}_{v, V', r} \coloneqq \Phi_{v, V', r} / \Psi_{v, V', r},
\label{eq:5}
\end{align}
where
\begin{align*}
\Phi_{v, V', r} &= \frac{1}{r} \sum_{k=1}^r \frac{f_v(X_k) \, \bm{1}_{\{v_{X_k} \in V'\}}}{d_{X_k}}, \\
\Psi_{v, V', r} &= \frac{1}{r} \sum_{k=1}^r \frac{\bm{1}_{\{v_{X_k} \in V'\}}}{d_{X_k}},
\end{align*}
where $\bm{1}_{\{\text{cond}\}}$ is an indicator function that returns 1 if the condition `cond' holds true and returns 0 otherwise.

We discuss the convergence of the estimators $\hat{\mu}_{v, r}$ and $\hat{\mu}_{v, V', r}$ as $r$ goes to infinity.
When we define $g((i, \alpha)) = f_v(i) / d_i$, it holds true that
\begin{align*}
\Phi_{v, r} &= \frac{1}{r} \sum_{k=1}^r g((X_k, Y_k)) \\
&\to \mathbb{E}_{\bm{\pi}_{\mathcal{E}}}[\Phi_{v, r}] \coloneq \sum_{(i, \alpha) \in \mathcal{E}} \frac{f_v(i)}{d_i} \pi_{\mathcal{E}, (i, \alpha)} = \sum_{i=1}^n \sum_{\alpha=1}^m \frac{f_v(i)}{d_i} \frac{b_{i, \alpha}}{D} =
\sum_{i=1}^n \frac{f_v(i)}{D}
\end{align*}
almost surely\ as $r \to \infty$ because of Theorem \ref{theorem:2}. 
When we define $g((i, \alpha)) = 1 / d_i$, it holds true that
\begin{align*}
\Psi_{v, r} &= \frac{1}{r} \sum_{k=1}^r g((X_k, Y_k)) \to \mathbb{E}_{\bm{\pi}_{\mathcal{E}}}[\Psi_{v, r}] \coloneq \sum_{(i, \alpha) \in \mathcal{E}} \frac{1}{d_i} \pi_{\mathcal{E}, (i, \alpha)} = \frac{n} {D}
\end{align*}
almost surely\ as $r \to \infty$ because of Theorem \ref{theorem:2}.
Therefore, the estimator $\hat{\mu}_{v, r}$ asymptotically converges to $\mu_{v, r}$ almost surely as $r \to \infty$ because 
\begin{align*}
\mu_{v, r} = \frac{\mathbb{E}_{\bm{\pi}_{\mathcal{E}}}[\Phi_{v, r}]}{\mathbb{E}_{\bm{\pi}_{\mathcal{E}}}[\Psi_{v, r}]}.
\end{align*}
Similarly, the estimator $\hat{\mu}_{v, V', r}$ asymptotically converges to $\mu_{v, V', r}$ almost surely as $r \to \infty$ because it holds true that $\Phi_{v, V', r} \to \mathbb{E}_{\bm{\pi}_{\mathcal{E}}}[\Phi_{v, V', r}] \coloneq \sum_{i=1, v_i \in V'}^n f_v(i) / D$ almost surely as $r \to \infty$, $\Psi_{v, V', r} \to \mathbb{E}_{\bm{\pi}_{\mathcal{E}}}[\Psi_{v, V', r}] \coloneq |V'| / D$ almost surely as $r \to \infty$, and $\mu_{v, V', r} = \mathbb{E}_{\bm{\pi}_{\mathcal{E}}}[\Phi_{v, V', r}] / \mathbb{E}_{\bm{\pi}_{\mathcal{E}}}[\Psi_{v, V', r}]$.

\subsubsection{Hyperedge properties}
We also estimate the hyperedge's property of interest in $H$ defined as $\mu_{e} \coloneqq \sum_{\alpha=1}^m f_e(\alpha) / m$, where $f_e: \{1, \ldots, m\} \to \mathbb{R}$ is a function that returns the feature of interest of hyperedge $e_{\alpha} \in E$.
As with the case of the node's property, one can specify the hyperedge's feature of interest by choosing $f_e$ appropriately.
For example, if $f_e(\alpha) = s_{\alpha}$, then $\mu_{e}$ is the average size of a hyperedge in $H$ \cite{luo2024}. 
If $f_e(\alpha)$ returns 1 if the hyperedge $e_{\alpha}$ has the given size $s$ and returns 0 otherwise, then $\mu_{e}$ is the probability that a hyperedge has size $s$; summing these probabilities over all possible s values yields an estimator for the probability distribution of the hyperedge size \cite{luo2024}.
We define an estimator of $\mu_{e}$ as
\begin{align}
\hat{\mu}_{e, r} \coloneqq \Phi_{e, r} / \Psi_{e, r},
\label{eq:6}
\end{align}
where
\begin{align*}
\Phi_{e, r} &= \frac{1}{r} \sum_{k=1}^r \frac{f_e(Y_k)}{s_{Y_k}}, \\
\Psi_{e, r} &= \frac{1}{r} \sum_{k=1}^r \frac{1}{s_{Y_k}}.
\end{align*}
Similar to the case of node, this estimator can be extended to the property of interest for the hyperedges in a subset $E' \subset E$ (e.g., the average size of a hyperedge in $E'$).
Let us estimate $\mu_{e, E'} \coloneqq \sum_{\alpha=1, e_{\alpha} \in E'}^m f_e(\alpha) / |E'|$.
We define an estimator of $\mu_{e, E'}$
\begin{align}
\hat{\mu}_{e, E', r} \coloneqq \Phi_{e, E', r} / \Psi_{e, E', r},
\label{eq:7}
\end{align}
where
\begin{align*}
\Phi_{e, E', r} &= \frac{1}{r} \sum_{k=1}^r \frac{f_e(Y_k) \, \bm{1}_{\{e_{Y_k} \in E'\}}}{s_{Y_k}}, \\
\Psi_{e, E', r} &= \frac{1}{r} \sum_{k=1}^r \frac{\bm{1}_{\{e_{Y_k} \in E'\}}}{s_{Y_k}}.
\end{align*}

We discuss the convergence of the estimators $\hat{\mu}_{e, r}$ and $\hat{\mu}_{e, E', r}$ as $r$ goes to infinity.
When we define $g((i, \alpha)) = f_e(\alpha) / s_{\alpha}$, it holds true that
\begin{align*}
\Phi_{e, r} &= \frac{1}{r} \sum_{k=1}^r g((X_k, Y_k)) \to \mathbb{E}_{\bm{\pi}_{\mathcal{E}}}[\Phi_{e, r}] \coloneq \sum_{i=1}^n \sum_{\alpha=1}^m \frac{f_e(\alpha)}{s_{\alpha}} \frac{b_{i, \alpha}}{D} = \sum_{\alpha=1}^m \frac{f_e(\alpha)}{D}
\end{align*}
almost surely\ as $r \to \infty$ because of Theorem \ref{theorem:2}.
When we define $g((i, \alpha)) = 1 / s_{\alpha}$, it holds true that
\begin{align*}
\Psi_{e, r} &= \frac{1}{r} \sum_{k=1}^r g((X_k, Y_k)) \to \mathbb{E}_{\bm{\pi}_{\mathcal{E}}}[\Psi_{e, r}] \coloneq \sum_{i=1}^n \sum_{\alpha=1}^m \frac{1}{s_{\alpha}} \frac{b_{i, \alpha}}{D} = \frac{m}{D}
\end{align*}
almost surely as $r \to \infty$ because of Theorem \ref{theorem:2}.
Therefore, the estimator $\hat{\mu}_{e, r}$ asymptotically converges to $\mu_{e, r}$ almost surely as $r \to \infty$ because 
\begin{align*}
\mu_{e, r} = \frac{\mathbb{E}_{\bm{\pi}_{\mathcal{E}}}[\Phi_{e, r}]}{\mathbb{E}_{\bm{\pi}_{\mathcal{E}}}[\Psi_{e, r}]}.
\end{align*}
Similarly, the estimator $\hat{\mu}_{e, E', r}$ asymptotically converges to $\mu_{e, E', r}$ almost surely as $r \to \infty$ because it holds true that $\Phi_{e, E', r} \to \mathbb{E}_{\bm{\pi}_{\mathcal{E}}}[\Phi_{e, E', r}] \coloneq \sum_{\alpha=1, e_{\alpha} \in E'}^m f_e(\alpha) / D$ almost surely as $r \to \infty$, $\Psi_{e, E', r} \to \mathbb{E}_{\bm{\pi}_{\mathcal{E}}}[\Psi_{e, E', r}] \coloneq |E'| / D$ almost surely as $r \to \infty$, and $\mu_{e, E', r} = \mathbb{E}_{\bm{\pi}_{\mathcal{E}}}[\Phi_{e, E', r}] / \mathbb{E}_{\bm{\pi}_{\mathcal{E}}}[\Psi_{e, E', r}]$.

\section{Results} \label{section:3}

\subsection{Simulation results in empirical hypergraphs}
\label{section:3.1}

\begin{table*}[t]
\caption{Empirical hypergraphs. $n$: number of nodes. $m$: number of hyperedges. $\bar{d}$: average degree of a node. $d_{\text{max}}$: maximum degree of the node. $P(d_i = 1)$: proportion of nodes with degree 1. $\bar{s}$: average size of a hyperedge}. $s_{\text{max}}$: maximum size of the hyperedge.
\begin{center}
\begin{tabular}{| A | B | B | C | D | E | C | F |} \hline 
Dataset & $n$ & $m$ & $\bar{d}$ & $d_{\text{max}}$ & $P(d_i=1)$ & $\bar{s}$ & $s_{\text{max}}$ \rule[0pt]{-5pt}{11pt} \\ \hline
MAG-geology & 1,088,242 & 1,013,444 & 3.5 & 1,030 & 0.592 & 3.8 & 283 \\
MAG-history & 373,713 & 248,754 & 2.0 & 1,574 & 0.685 & 3.1 & 919 \\ 
DBLP & 1,748,508 & 2,908,856 & 5.4 & 1,271 & 0.480 & 3.3 & 273 \\ 
Amazon & 2,268,192 & 4,285,330 & 32.2 & 28,973 & 0.0006 & 17.1 & 9,350 \\ 
stack-overflow & 2,524,925 & 8,790,363 & 9.1 & 32,797 & 0.454 & 2.6 & 67 \\ 
\hline
\end{tabular}
\label{table:1}
\end{center}
\end{table*}

We use five empirical hypergraphs.
The MAG-geology and MAG-history hypergraphs are composed of authors (i.e., nodes) and publications co-written by authors (i.e., hyperedges) indexed in the Microsoft Academic Graph (MAG) database \cite{amburg2020,sinha2015,benson2018}.
The DBLP hypergraph consists of authors (i.e., nodes) and publications co-written by authors (i.e., hyperedges) indexed in the DBLP database \cite{benson2018}.
The Amazon hypergraph consists of products (i.e., nodes) and product reviews generated by users on Amazon (i.e., hyperedges).
The stack-overflow hypergraph consists of users (i.e., nodes) in the Stack Overflow--a question-and-answer website for computer programmers--and threads in which users participated (i.e., hyperedges).
We use the largest connected component of each hypergraph.
Table \ref{table:1} shows the basic properties of these hypergraphs.
We present the results for the DBLP hypergraph in the remainder of this subsection; see Supplementary Section~S1 for the results in the other hypergraphs.

\begin{figure}[t]
    \centering
    \includegraphics[width=1.0\linewidth]{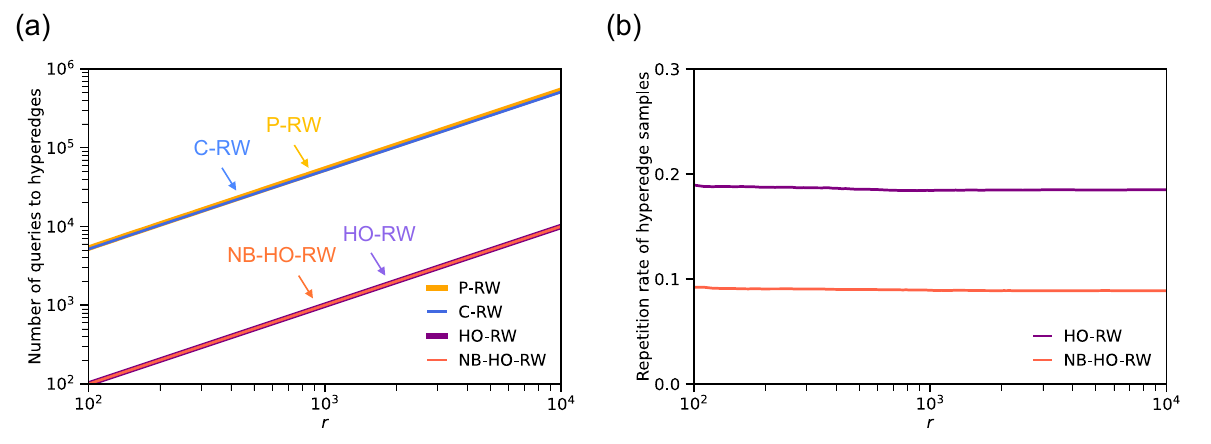}\\
    \caption{Comparison of random walks from a sampling perspective in the DBLP hypergraph. (a) The number of queries to hyperedges. The curves for P-RW and C-RW heavily overlap with each other, and those for HO-RW and NB-HO-RW completely overlap with each other. (b) The repetition rate of hyperedge samples.}
\label{fig:2}
\end{figure}

A random walk that requires fewer queries is preferable since practical scenarios often impose a limit on the number of queries used to collect data \cite{gjoka2011, hardiman2013}.
We compare four random walks (P-RW, C-RW, HO-RW, and NB-HO-RW) in terms of their query requirements.
First, we independently execute each random walk for a given length $r$.  
Second, we compute the number of queries to hyperedges (including duplicates) for each random walk.
We omit the number of queries to nodes (including duplicates) for each random walk because it is the same for all four random walks (i.e., $r$ queries) for any $r$.

Figure \ref{fig:2}(a) shows the average number of queries to hyperedges over $10^3$ independent runs of each random walk as a function of $r$.
For any $r \in [10^2, 10^4]$, we observe that P-RW and C-RW generate a considerably larger number of queries compared to the others, while HO-RW and NB-HO-RW generate the same number of queries, as expected.

A random walk with fewer repeated samples may yield better estimation results \cite{lee2012, li2015}.
We compare the repetition rate of hyperedge samples, defined as $|\{(Y_{k}, Y_{k+1})\ |\ 1 \leq k \leq r-1 \land Y_k = Y_{k+1}\} | / (r-1)$, between HO-RW and NB-HO-RW.
We omit the results for the repetition rate of node samples, defined as $|\{(X_{k}, X_{k+1})\ |\ 1 \leq k \leq r-1 \land X_k = X_{k+1}\} | / (r-1)$, because it is zero for both the random walks for any $r$.

Figure \ref{fig:2}(b) shows the average repetition rate of hyperedge samples over $10^3$ independent runs of each random walk as a function of $r$.
As expected, we confirm that NB-HO-RW consistently achieves a lower repetition rate of hyperedge samples than HO-RW for any $r \in [10^2, 10^4]$.

\begin{figure}[t]
    \centering
    \includegraphics[width=1.0\linewidth]{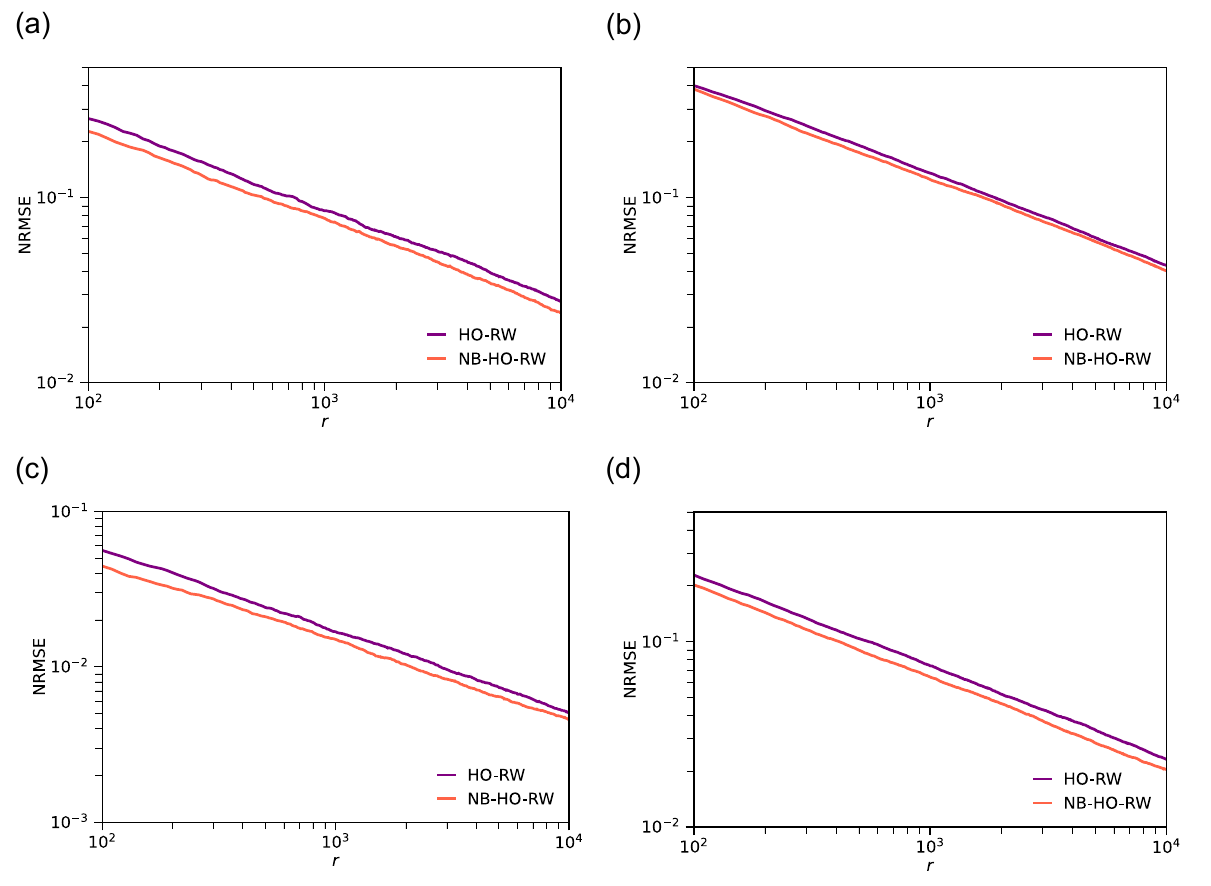}\\
    \caption{NRMSE of the estimators of node and hyperedge properties as a function of $r$ in the DBLP hypergraph. (a) Average degree of a node. (b) Probability distribution of the node degree. (c) Average size of a hyperedge}. (d) Probability distribution of the hyperedge size.
\label{fig:3}
\end{figure}

We now assess estimators of node and hyperedge properties based on HO-RW and NB-HO-RW.
To estimate specific properties, we define:
(i) $f_{v}(i) = d_i$ in Eq.~\eqref{eq:4} for estimating the average degree of a node; (ii) $f_{v}(i) = \bm{1}_{\{d_{i} = d\}}$ for each $d = 1, \ldots, m$ in Eq.~\eqref{eq:4} for estimating the probability distribution of the node degree;
(iii) $f_{e}(\alpha) = s_{\alpha}$ in Eq.~\eqref{eq:6} for estimating the average size of a hyperedge; and
(iv) $f_{e}(\alpha) = \bm{1}_{\{s_{\alpha} = s\}}$ for each $s = 2, \ldots, n$ in Eq.~\eqref{eq:6} for estimating the probability distribution of the hyperedge size.
In practice, an estimator should have both low bias in a single run and small variance across different runs.  
The normalized root mean squared error (NRMSE) evaluates both bias and variance \cite{ribeiro2010, lee2012, hardiman2013, chen2016, nakajima2023}.  
We define the NRMSE of an estimator as $\sqrt{\mathbb{E}[(\Delta(\hat{x}, x)^2]}$, where $x$ is the original value, $\hat{x}$ is the estimator, and $\Delta(\hat{x}, x)$ is a certain error metric between $x$ and $\hat{x}$ \cite{nakajima2023}.
We use the relative error to evaluate the estimators of the average degree of a node and the average size of a hyperedge. 
We use the $L^1$ distance to evaluate the estimators of the probability distributions of the node degree and the hyperedge size.
To estimate the NRMSE of each estimator, we independently perform HO-RW and NB-HO-RW of the given length $r$ each $10^3$ times.

Figure \ref{fig:3} shows the NRMSE of each estimator as a function of $r$.
NB-HO-RW slightly reduces the NRMSE of each estimator compared to HO-RW for any $r \in [10^2, 10^4]$.
For both HO-RW and NB-HO-RW, the NRMSE of each estimator decreases approximately in inverse proportion to the square root of $r$.
For NB-HO-RW with $r = 10^4$, the NRMSE reaches $2.4 \times 10^{-2}$ in Fig.~\ref{fig:3}(a), $4.0 \times 10^{-2}$ in Fig.~\ref{fig:3}(b), $4.6 \times 10^{-3}$ in Fig.~\ref{fig:3}(c), and $2.0 \times 10^{-2}$ in Fig.~\ref{fig:3}(d).
Notably, $r = 10^4$ remains considerably smaller than $n$ and $m$ in the DBLP hypergraph, specifically $r \approx 0.006 n$ and $r \approx 0.003 m$.

\begin{figure}[t]
    \centering
    \includegraphics[width=1.0\linewidth]{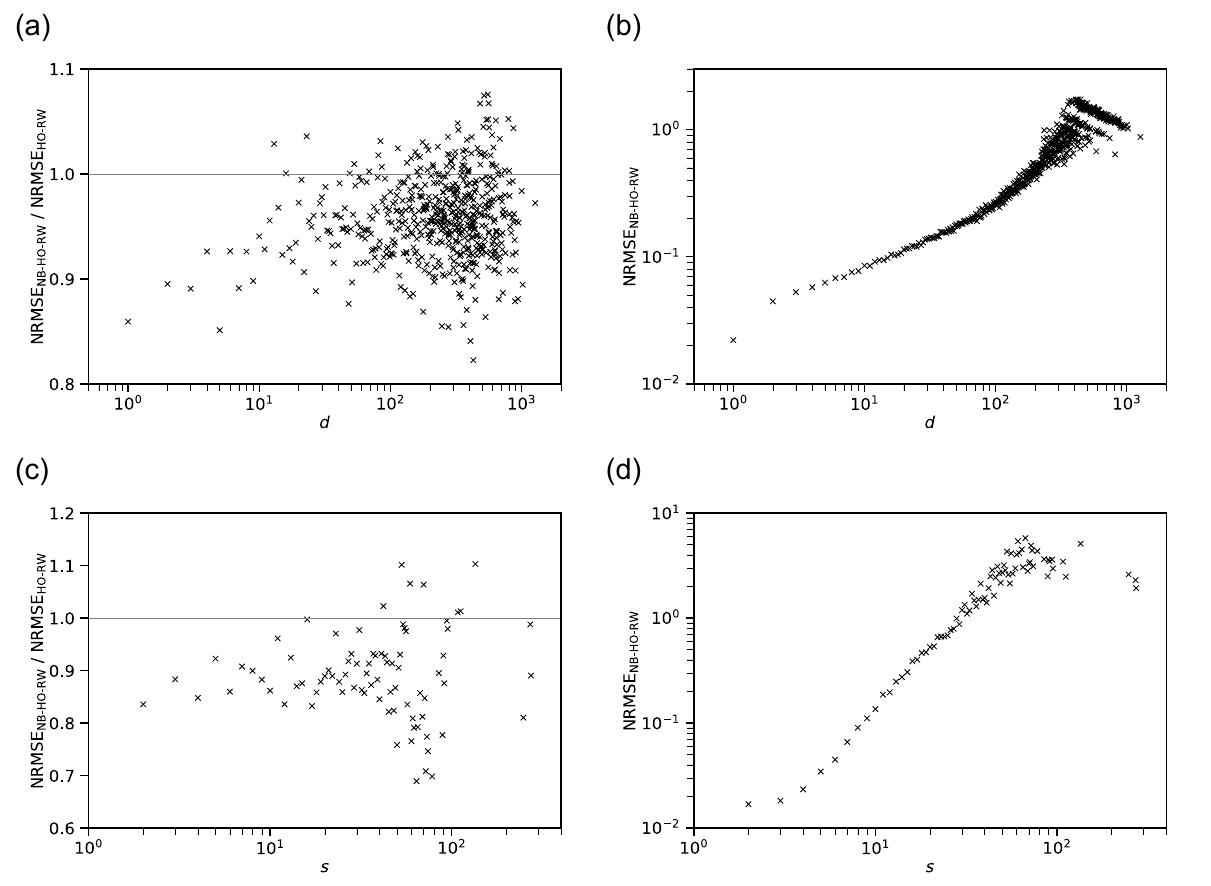}\\
    \caption{NRMSE of the estimators of node and hyperedge properties for $r = 10^4$ in the DBLP hypergraph. (a) and (b) Probability that a node has degree $d$ as a function of $d$. (c) and (d) Probability that a hyperedge has size $s$ as a function of $s$. $\text{NRMSE}_{\text{HO-RW}}$ and $\text{NRMSE}_{\text{NB-HO-RW}}$ indicate the NRMSEs for HO-RW and NB-HO-RW, respectively.}
\label{fig:4}
\end{figure}

We validate the estimators for the probability distributions of node degree and hyperedge size in more detail.  
Figure \ref{fig:4}(a) shows the ratio of the NRMSE of the estimator for the probability of nodes with degree $d$ for the NB-HO-RW to that for the HO-RW as a function of $d$.  
NB-HO-RW reduces the NRMSE compared to HO-RW (i.e., the ratio is less than 1) for many values of $d$.  
Figure \ref{fig:4}(b) shows the NRMSE for NB-HO-RW as a function of $d$.  
The NRMSE tends to increase with $d$, consistent with previous results for graphs \cite{lee2012}.  
Figures \ref{fig:4}(c) and \ref{fig:4}(d) show the results for the estimators of the probability distribution of hyperedge size.  
Similar to Figs.~\ref{fig:4}(a) and \ref{fig:4}(b), NB-HO-RW reduces the NRMSE compared to HO-RW for many values of $s$, and the NRMSE for NB-HO-RW tends to increase with $s$.

We obtained qualitatively the same results as for the DBLP hypergraph in the MAG-geology, MAG-history, and stack-overflow hypergraphs (see Supplementary Section S1 for details).
For the Amazon hypergraph, HO-RW and NB-HO-RW yield comparable NRMSE values for each estimator (see Supplementary Section S1 for details).  
This is because the repetition ratio is already sufficiently small for HO-RW due to the low proportion of nodes with degree 1 in the Amazon hypergraph (see Supplementary Section S1 for details; see also Table \ref{table:1}).  
Note that in NB-HO-RW, a walker at a node with degree 1 returns to the previous hyperedge.  
In many real-world hypergraphs, the node degree distribution has a heavy right tail, and the proportion of nodes with degree 1 is large \cite{do2020}.  
Thus, we conclude that estimators based on NB-HO-RW often yield better estimates than those based on HO-RW in large empirical hypergraphs.

\subsection{Sampling and estimation in OpenAlex} \label{section:3.2}

We showcase the application of NB-HO-RW to a hypergraph where the full snapshot is not at hand.
OpenAlex is an open database of publications with rich bibliographic information across various disciplines~\cite{priem2022}.
We focus on the underlying hypergraph consisting of the set of authors (i.e., nodes) indexed in the database and the set of publications co-authored by authors (i.e., hyperedges).
As of February 2025, OpenAlex indexed 102 million author IDs\footnote{\url{https://api.openalex.org/authors?search=} (accessed February 2025)} and 263 million publication IDs\footnote{\url{https://api.openalex.org/works?search=} (accessed February 2025)}.

The OpenAlex hypergraph satisfies the four assumptions made in our problem definition.
First, the set of publication IDs of author $u$ (i.e., those in which $u$ is an author) accessible via an API query: `https://api.openalex.org/works?filter=author.id:$\langle u$'s ID$\rangle$'.
When retrieving the publication list of author $u$ through the API, we excluded their sole-author publications.
Second, the set of author IDs for publication $z$ is available available via an API query: `https://api.openalex.org/works/$\langle z$'s ID$\rangle$'.
Third, the hypergraph is considered static during a short sampling period, as the database snapshot is updated approximately once per month in the database\footnote{\url{https://docs.openalex.org/download-all-data/openalex-snapshot} (accessed February 2025)}.
Fourth, we set the seed of NB-HO-RW as the author ID of `Federico Battiston', the first author of Ref.~\cite{battiston2020}.

We performed NB-HO-RW with a length of $r' = 4 \times 10^4$ on the hypergraph, with the sampling period between February 8 and February 12, 2025.
Note that $r$ is approximately 0.4\% of the total number of authors ($\approx$ 102 million) and 0.2\% of the total number of publications ($\approx$ 263 million).
We adhered to the daily API call limit of 100,000 requests per user per day as of February 2025\footnote{\url{https://docs.openalex.org/how-to-use-the-api/api-overview} (accessed February 2025)}.
Although the entire OpenAlex hypergraph may not be fully connected, it is unlikely that a random walk got stuck in an isolated small component of the original hypergraph because the sample sequence contains unique 34,944 authors and 34,329 publications.
To account for the dependence on the random walk seed \cite{gjoka2011}, we discarded the first $5 \times 10^3$ and used samples from the 5,001st to the $r'$-th step, where $5000 < r' \leq 40000$.

The API output for each publication $z$ in the sample sequence includes $z$'s primary research domain (`Health Sciences', `Life Sciences', `Physical Sciences', or `Social Sciences').
We classify a publication into a research domain based on its primary research domain.  
Using this classification, we assign each author $u$ in the sample sequence the most common research domains appearing in $u$'s publications \cite{huang2020, nakajima20232}.  
We define authors in a research domain as those whose most common research domains include that domain.

We estimate the average and complementary cumulative distribution function (CCDF) of author productivity (i.e., node degree) and team size (i.e., hyperedge size) both across all four research domains combined and within each domain separately.
To estimate the CCDF of author productivity, we define $f_{v}(i) = \bm{1}_{\{d_{i} \geq d\}}$ for each $d = 1, \ldots, m$ in Eqs.~\eqref{eq:4} and \eqref{eq:5}.
To estimate the CCDF of team size, we define $f_{e}(\alpha) = \bm{1}_{\{s_{\alpha} \geq s\}}$ for each $s = 2, \ldots, n$ in Eqs.~\eqref{eq:6} and \eqref{eq:7}.
For estimates by research domain, we define $V'$ as the set of authors in a given research domain in Eq.~\eqref{eq:5} and $E'$ as the set of publications in a given research domain in Eq.~\eqref{eq:7}.

\begin{figure}[t]
    \centering
    \includegraphics[width=1.0\linewidth]{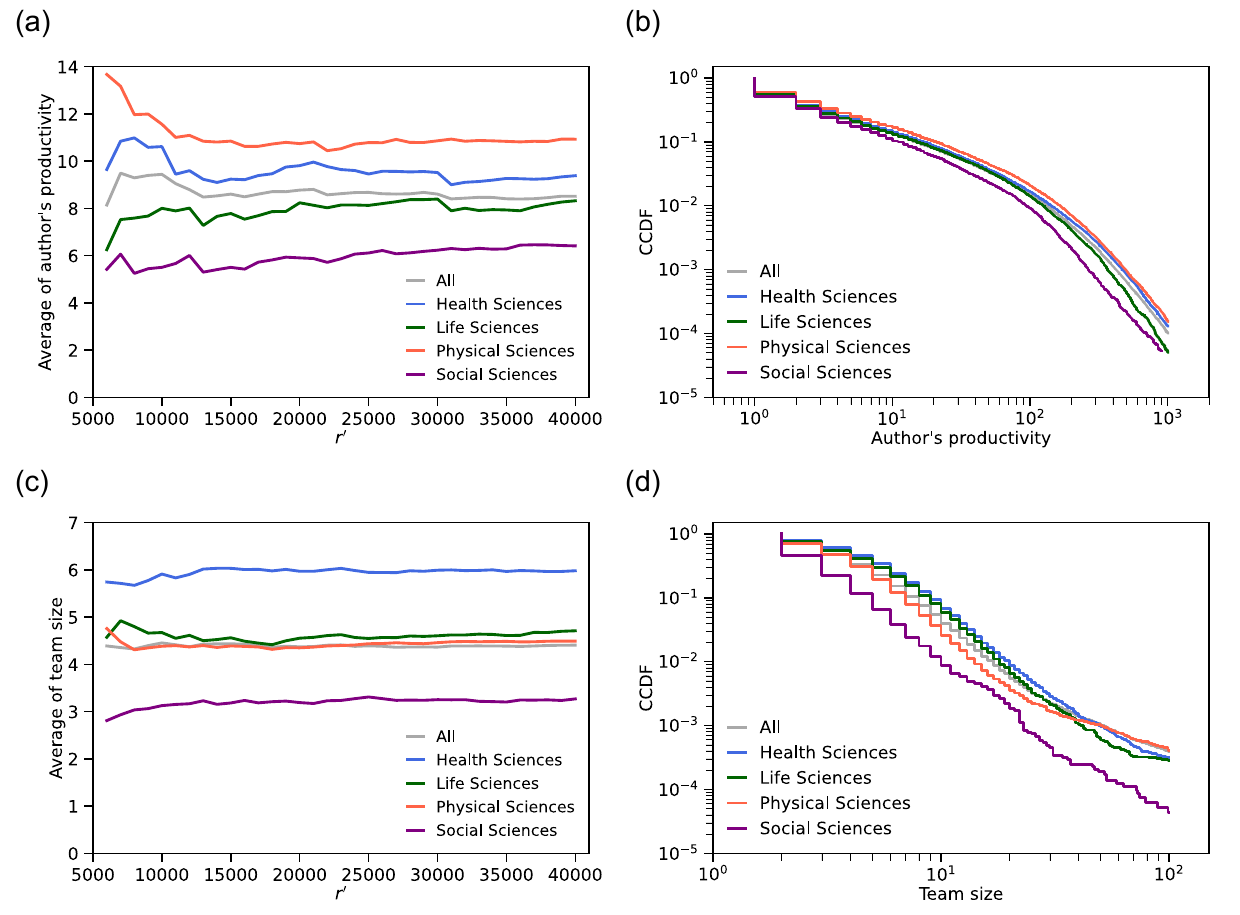}\\
    \caption{Estimation of properties of authors and publications indexed in the OpenAlex database across all research domains and by research domain. (a) Average productivity of authors. (b) Productivity distribution of authors. (c) Average team size of publications. (d) Size distribution of publications. We set $r' = 4.0 \times 10^4$ in (b) and (d). }
\label{fig:5}
\end{figure}

Figure \ref{fig:5}(a) shows the estimated average productivity of authors across all domains and within each research domain.  
The estimates stabilize at approximately $r' = 4 \times 10^4$.  
Compared to the overall average, author productivity is higher in the Health and Physical Sciences, similar in the Life Sciences, and lower in the Social Sciences.  
Figure \ref{fig:5}(b) shows the estimated distribution of author productivity across all domains and by research domain, with $r' = 4 \times 10^4$.  
The probability of authors having published $d$ papers declines more rapidly in the Social Sciences than in other domains as $d$ increases.

Figure \ref{fig:5}(c) shows the estimated average team size of publications across all domains and within each research domain as a function of $r'$.  
The estimates converge to stable values around $r' = 4 \times 10^4$.  
Compared to the overall average, the average team size is higher in the Health Sciences, similar in the Physical and Life Sciences, and lower in the Social Sciences.  
This result is partially consistent with previous findings on team science \cite{wuchty2007, lariviere2015}.  
Figure \ref{fig:5}(d) shows the estimated distribution of team size across all domains and by research domain, with $r' = 4 \times 10^4$.  
The probability of publications with $s$ authors declines more rapidly in the Social Sciences than in other domains as $s$ increases.

\begin{figure}[t]
    \centering
    \includegraphics[width=1\linewidth]{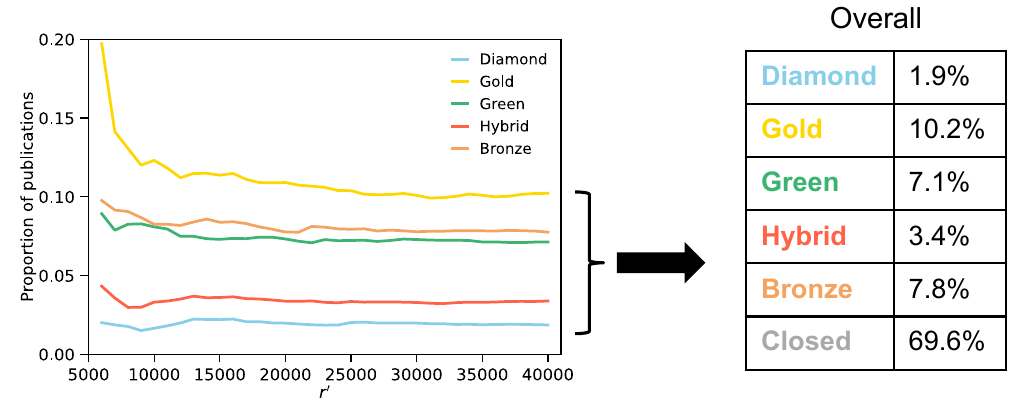}\\
    \caption{Estimation of the OA state for publications indexed in OpenAlex. Estimated proportion of each OA publication category as a function of the number of samples used (left). The estimates obtained for $r' = 4 \times 10^4$ (right).}
\label{fig:6}
\end{figure}

There is growing interest in Open Access (OA) to scientific literature and a need to assess the prevalence and characteristics of OA publications \cite{piwowar2018}.  
The API output for each publication $z$ in the sample sequence includes $z$'s OA status, which is one of the following\footnote{\url{https://help.openalex.org/hc/en-us/articles/24347035046295-Open-Access-OA} (Accessed February 2025)}:  
\begin{itemize}  
\item Diamond: Published in a fully OA journal without article processing charges.  
\item Gold: Published in a fully OA journal.  
\item Green: Toll-access on the publisher’s landing page, with a free copy available in an OA repository.  
\item Hybrid: Free under an open license in a toll-access journal.  
\item Bronze: Free to read on the publisher’s landing page without an identifiable license.  
\item Closed: All other articles.  
\end{itemize}  
We classify publications with Diamond, Gold, Green, Hybrid, or Bronze OA status as OA publications.  
We estimate the composition of OA statuses across all research domains and within each research domain.  
To this end, we define $f_{e}(\alpha) = \bm{1}_{\{e_{\alpha}\text{ has OA status } o\}}$ for each $o \in \{\text{Diamond}, \text{Gold}, \text{Green}, \text{Hybrid}, \text{Bronze}, \text{Closed}\}$ in Eqs.~\eqref{eq:6} and \eqref{eq:7}.

\begin{figure}[t]
    \centering
    \includegraphics[width=1\linewidth]{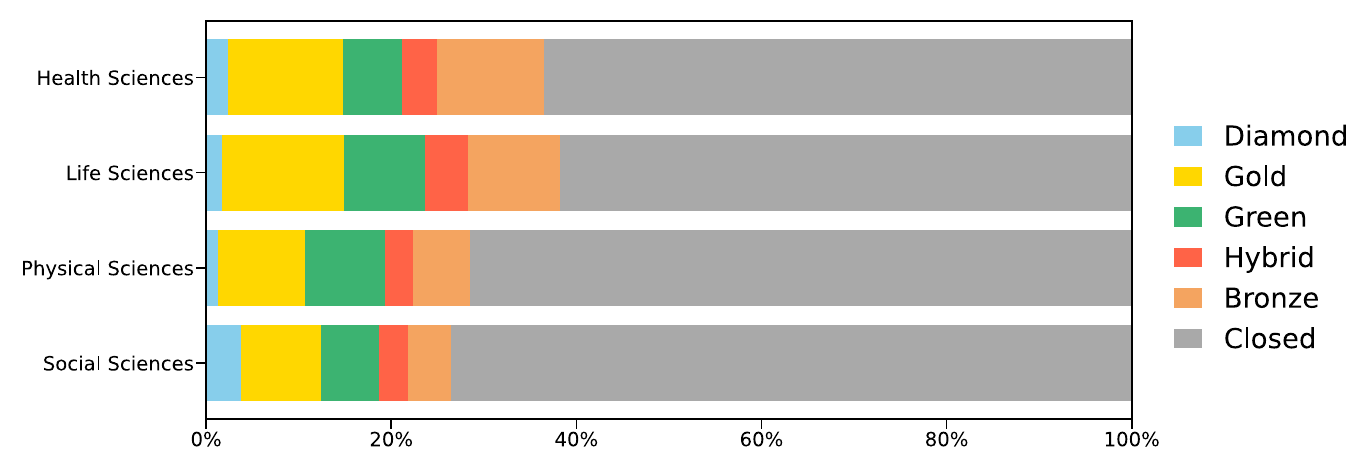}\\
    \caption{Estimated proportion of publications with each OA status by research domain, with $r' = 4.0 \times 10^4$.}
\label{fig:7}
\end{figure}

Figure \ref{fig:6} (left) shows the estimated proportion of each OA status, excluding "Closed," as a function of $r'$.  
The estimates stabilize around $r' = 4 \times 10^4$.  
Figure \ref{fig:6} (right) shows the estimated compositional ratio of OA statuses for publications, with $r' = 4 \times 10^4$.  
Gold and Bronze OA statuses dominate OA publications, which is largely consistent with previous findings \cite{piwowar2018}.  

Figure \ref{fig:7} shows the estimated compositional ratio of OA statuses by research domain, with $r' = 4 \times 10^4$.  
OpenAlex indexes a relatively large number of Diamond OA journals \cite{simard2024}.  
Our estimates indicate that the share of Diamond OA is higher in the Social Sciences than in other domains.  
The proportions of Gold and Bronze OA are higher in the Health and Life Sciences, while Green OA is more common in the Life and Physical Sciences.  
These findings are largely consistent with previous results \cite{piwowar2018}.  
The proportion of Hybrid OA is comparable across the four domains.

\begin{figure}[t]
    \centering
    \includegraphics[width=1\linewidth]{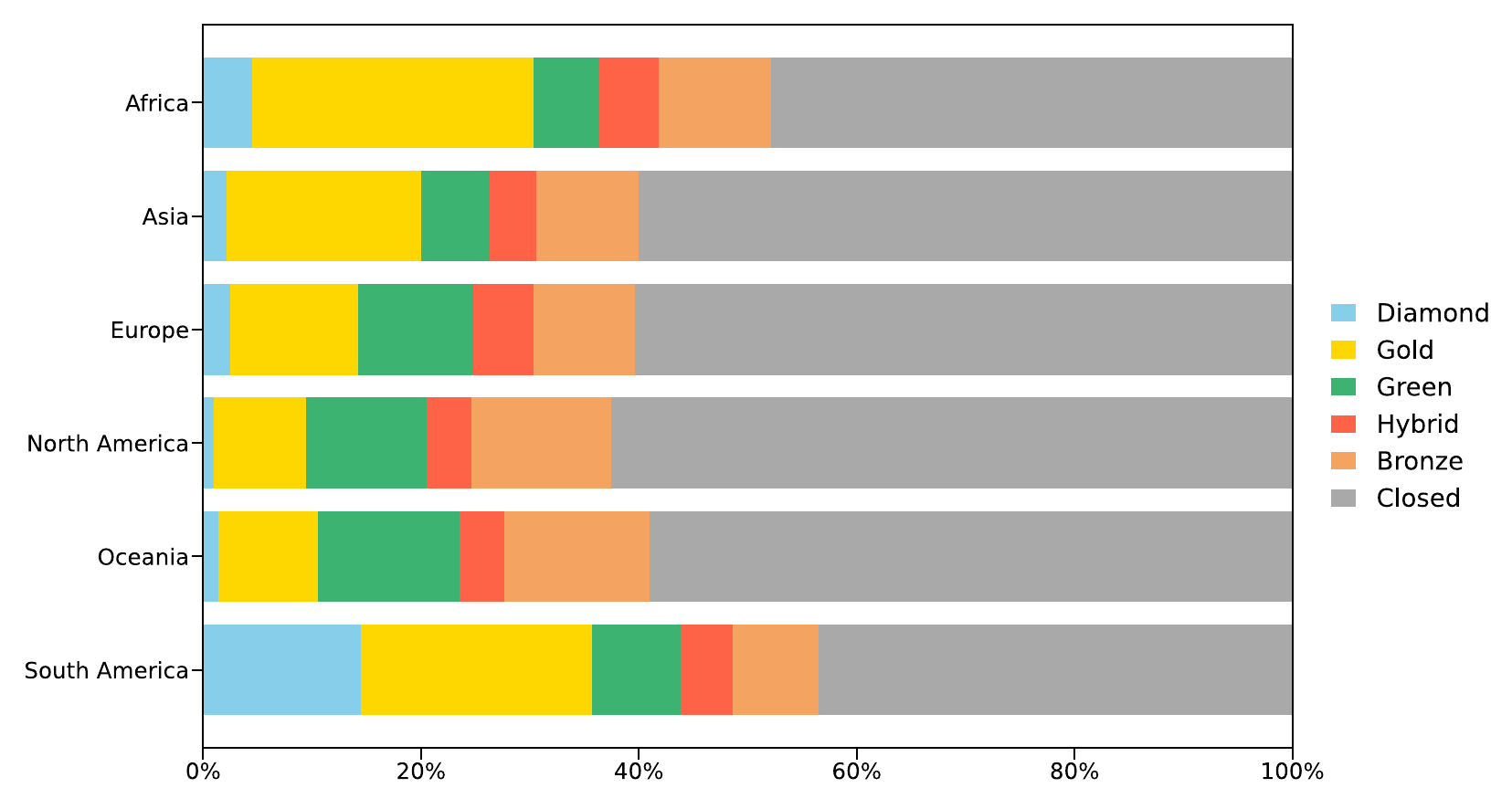}\\
    \caption{Estimated proportion of publications with each OA status by geographic region, with $r' = 4.0 \times 10^4$.}
\label{fig:8}
\end{figure}

We also compare the adoption of OA publications across geographic regions.  
The API output for each publication in the sample sequence includes the country codes of the authors' affiliations.  
Using the `country\_converter` library, we obtained the continent codes of the authors' affiliations for each publication.  
We focus on the following continents (hereafter referred to as regions): `Africa', `Asia', `Europe', `North America', `Oceania', and `South America'.  
We classify a publication as belonging to a region if the continent codes of the authors' affiliations include that region.  
To obtain estimates by geographic region, we define $E'$ as the set of publications in a given geographic region in Eq.~\eqref{eq:7}.

Figure \ref{fig:8} shows the estimated compositional ratio of OA statuses for publications by geographic region, with $r' = 4 \times 10^4$.  
We observe the following trends.  
The share of OA publications is higher in Africa and South America than in other regions, consistent with previous findings \cite{demeter2021, basson2022, simard2022}.  
The proportion of Diamond OA is considerably higher in South America than in other regions, aligning with prior findings \cite{bosman2021}.  
Gold OA publications are more widely adopted in Africa, Asia, and South America, consistent with previous research \cite{simard2022, basson2022}.  
Green OA is more prevalent in Europe, North America, and Oceania, in line with earlier studies \cite{basson2022, simard2022}.  
The proportion of Hybrid OA is similar across all six regions.  
Finally, the share of Bronze OA is relatively high in North America and Oceania, consistent with previous findings \cite{simard2022, basson2022}.

\section{Conclusions} \label{section:4}

In this study, we focused on sampling nodes and hyperedges and estimating node and hyperedge properties using random walks on hypergraphs. 
Within a general framework for random walks on hypergraphs, we first described the differences between three existing random walks--P-RW \cite{zhou2006, battiston2020, zhang2023, traversa2024}, C-RW \cite{carletti2020}, and HO-RW \cite{banerjee2021, luo2024, traversa2024}--in terms of transition probabilities between nodes and the number of queries. 
We then extended HO-RW to NB-HO-RW, a random walk that avoids backtracking. 
We discussed the convergence of estimators based on NB-HO-RW.
Our numerical experiments show that NB-HO-RW yields lower NRMSE values in the estimation of certain node degree and hyperedge size properties compared to HO-RW. 
Estimators of various graph properties based on random walks have been extensively studied (e.g., \cite{gjoka2011, ribeiro2010, hardiman2013, wang2014, chen2016}). 
Future work is expected to extend these estimators to other hypergraph properties, such as node degree correlation \cite{latapy2008, nakajima2022}, node redundancy \cite{latapy2008, nakajima2022}, and hyperedge overlap \cite{lee2021, malizia2025}, using NB-HO-RW.
In addition, we did not address the convergence speed or sample complexity of estimators derived from NB-HO-RW, leaving this as an important topic for future theoretical investigation.

We applied NB-HO-RW to the OpenAlex database, where full hypergraph data is not readily available.  
We provided estimates for the distributions of author productivity, team size, and the prevalence of OA publications.  
Additionally, we computed these estimates for subsets of publications and authors, such as those from specific research domains or geographic regions.  
Although the number of samples collected by NB-HO-RW is small relative to the size of the OpenAlex hypergraph, we found that these estimates largely align with previous results from other bibliographic data samples.  
We acknowledge that our estimates are based on only one realization of the non-backtracking higher-order random walk on the OpenAlex hypergraph. 
Therefore, although our results largely align with previous findings in bibliographic analyses, our conclusions may vary across different realizations.
In cases where full access to resources like the Web of Science is unavailable, technical challenges in sampling publication data may arise when assessing the global research landscape, such as the prevalence of OA publications \cite{piwowar2018, simard2022}.
In response to such concerns, the Institute for Scientific Information recently examined the impact of publication sample size on citation statistics \cite{rogers2020}.  
While the OpenAlex database is fully open \cite{priem2022}, storing the entire publication dataset on a local machine requires substantial computational resources.  
We demonstrated that random walks can be useful in such cases for sampling publication data from large bibliographic databases and estimating the global research landscape.

\section*{Acknowledgments}

This work was supported in part by the JST ASPIRE Grant Number JPMJAP2328, JST ACT-X Grant Number JPMJAX24CI, the Nakajima Foundation, and TMU local 5G research support.

\appendix

\def\thesection{Appendix \Alph{section}}

\section{Proof of Lemma 1} \label{appendix:a}

First, for any $(i, \alpha) \in \mathcal{E}$ and any $(j, \beta) \in \mathcal{E}$, we get
\begin{align}
&U_{(i, \alpha), (j, \beta)} \notag \\
&\coloneq P(X_{k+1} = j \land Y_{k+1} = \beta\ |\ X_k = i \land Y_k = \alpha) \notag \\
&= P(Y_{k+1} = \beta\ |\ X_{k+1} = j \land X_k = i \land Y_k = \alpha) \, P(X_{k+1} = j\ |\ X_k = i \land Y_k = \alpha) \label{eq:a1} \\
&= P(Y_{k+1} = \beta\ |\ X_{k+1} = j \land Y_k = \alpha) \, P(X_{k+1} = j\ |\ X_k = i \land Y_k = \alpha), \label{eq:a2}
\end{align}
where Eq.~\eqref{eq:a1} is obtained by the definition of the conditional probability and Eq.~\eqref{eq:a2} is obtained by the definition of NB-HO-RW.
By the definition of NB-HO-RW, for any $(i, \alpha) \in \mathcal{E}$ and any $(j, \beta) \in \mathcal{E}$, we get
\begin{align}
P(Y_{k+1} = \beta\ |\ X_{k+1} = j \land Y_k = \alpha) &= 
\begin{dcases}
\bm{1}_{\{\alpha = \beta\}} & \text{if $d_j = 1$}, \\
\displaystyle \frac{\bm{1}_{\{\alpha \neq \beta\}}}{d_j-1} & \text{if $d_j > 1$},
\end{dcases} \label{eq:a3} \\
P(X_{k+1} = j\ |\ X_k = i \land Y_k = \alpha) &= \frac{b_{j, \alpha} \bm{1}_{\{i \neq j\}}}{s_{\alpha} - 1}, \label{eq:a4}
\end{align}
where recall that $\bm{1}_{\{\text{cond}\}}$ is an indicator function that returns 1 if the condition `cond' holds true and returns 0 otherwise.
Because of Eqs.~\eqref{eq:a2}--\eqref{eq:a4}, for any $(i, \alpha) \in \mathcal{E}$ and any $(j, \beta) \in \mathcal{E}$, we obtain 
\begin{align*}
U_{(i, \alpha), (j, \beta)} =
\displaystyle \begin{dcases}
\frac{b_{j, \alpha} \bm{1}_{\{i \neq j \land \alpha = \beta\}}}{s_{\alpha} - 1}    &   \text{if $d_j = 1$}, \\
\frac{b_{j, \alpha} \bm{1}_{\{i \neq j \land \alpha \neq \beta\}}}{(d_j-1)(s_{\alpha} - 1)} &   \text{if $d_j > 1$}.
\end{dcases}
\end{align*}

\section{Proof of Lemma 2} \label{appendix:b}

The stationary distribution $\bm{\pi}_{\mathcal{E}} \coloneq (\pi_{\mathcal{E}, (i, \alpha)})_{(i, \alpha) \in \mathcal{E}}$ uniquely exists because the chain is irreducible and aperiodic.
We will prove that a $D$-dimensional row vector, $(1/D)_{(i, \alpha) \in \mathcal{E}}$, is the stationary distribution of the chain, where recall that $D \coloneq |\mathcal{E}| = \sum_{i=1}^n d_i = \sum_{\alpha=1}^m s_{\alpha}$.
To this end, it is sufficient to show that the sum of any column vector of the matrix $\bm{U}$ is equal to 1; note that this is nontrivial for the transition probability matrix of a Markov chain.
First, for any $(j, \beta) \in \mathcal{E}$ such that $d_j = 1$, we get
\begin{align}
\sum_{(i, \alpha) \in \mathcal{E}} U_{(i, \alpha), (j, \beta)} &= \sum_{(i, \alpha) \in \mathcal{E}} \frac{b_{j, \alpha} \, \bm{1}_{\{i \neq j \land \alpha = \beta\}}}{s_{\alpha} - 1} \notag \\
&= \sum_{i=1}^n \sum_{\alpha=1}^m \frac{b_{i, \alpha} \, b_{j, \alpha} \, \bm{1}_{\{i \neq j \land \alpha = \beta\}}}{s_{\alpha} - 1} \notag\\
&= \sum_{\substack{i=1 \\ i \neq j}}^n \frac{b_{i, \beta} \, b_{j, \beta}}{s_{\beta} - 1} \notag\\
&= 1, \label{eq:b5}
\end{align}
where Eq.~\eqref{eq:b5} holds true because of $(j, \beta) \in \mathcal{E}$, i.e., $b_{j, \beta} = 1$.
Second, for any $(j, \beta) \in \mathcal{E}$ such that $d_j > 1$, we get
\begin{align}
\sum_{(i, \alpha) \in \mathcal{E}} U_{(i, \alpha), (j, \beta)} &= \sum_{(i, \alpha) \in \mathcal{E}} \frac{b_{j, \alpha} \, \bm{1}_{\{i \neq j \land \alpha \neq \beta\}}}{(d_j-1)(s_{\alpha} - 1)} \notag \\
&= \frac{1}{d_j-1} \sum_{i=1}^n \sum_{\alpha=1}^m \frac{b_{i, \alpha} b_{j, \alpha} \, \bm{1}_{\{i \neq j \land \alpha \neq \beta\}}}{s_{\alpha} - 1} \notag \\
&= \frac{1}{d_j-1} \sum_{\substack{i=1 \\ i \neq j}}^n \sum_{\substack{\alpha=1 \\ \alpha \neq \beta}}^m \frac{b_{i, \alpha} \, b_{j, \alpha}}{s_{\alpha} - 1} \notag \\
&= \frac{1}{d_j-1} \sum_{\substack{\alpha=1 \\ \alpha \neq \beta}}^m b_{j, \alpha} \sum_{\substack{i=1 \\ i \neq j}}^n \frac{b_{i, \alpha}}{s_{\alpha} - 1} \notag \\
&= \frac{1}{d_j-1} \sum_{\substack{\alpha=1 \\ \alpha \neq \beta}}^m b_{j, \alpha} \label{eq:b6} \\
&= 1, \label{eq:b7}
\end{align}
where Eq.~\eqref{eq:b6} holds true because it holds true that $\sum_{i=1,\ i \neq j}^n b_{i, \alpha} = s_{\alpha}-1$ for any $e_{\alpha} \in \mathcal{E} \backslash \{e_{\beta}\}$ such that $b_{j, \alpha} = 1$, and Eq.~\eqref{eq:b7} holds true because of $b_{j, \beta} = 1$.
Therefore, the stationary distribution is given by $\bm{\pi}_{\mathcal{E}} = (1 / D)_{(i, \alpha) \in \mathcal{E}}$, satisfying $\bm{\pi}_{\mathcal{E}}^\top = \bm{\pi}_{\mathcal{E}}^\top \bm{U}$.

\newpage

\begin{center}
\vspace*{12pt}
{\Large Supplementary Materials for:\\
\vspace{12pt}
Sampling nodes and hyperedges via random walks on large hypergraphs
}
\vspace{12pt} \\
\end{center}

\setcounter{figure}{0}
\setcounter{table}{0}
\setcounter{section}{0}

\renewcommand{\thesection}{S\arabic{section}}
\renewcommand{\thefigure}{S\arabic{figure}}
\renewcommand{\thetable}{S\arabic{table}}
\renewcommand{\theequation}{S\arabic{equation}}

\begin{center}
\author{Kazuki Nakajima, Masanao Kodakari, and Masaki Aida}
\vspace{24pt} \\
\end{center}

\section{Simulation results for other hypergraphs}

\begin{figure}[t]
    \centering
    \includegraphics[width=1.0\linewidth]{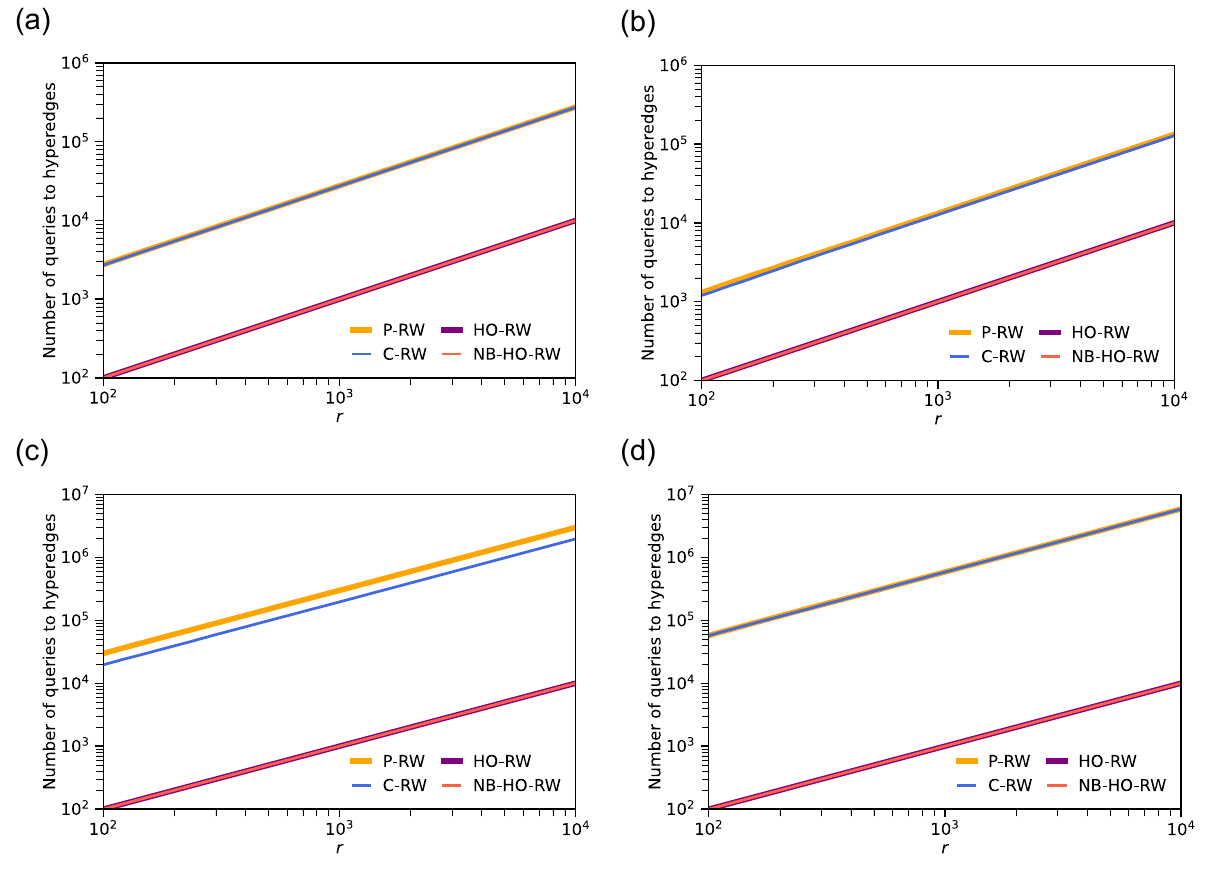}\\
    \caption{Comparison of the number of queries to hyperedges. (a) MAG-geology hypergraph. (b) MAG-history hypergraph. (c) Amazon hypergraph. (d) stack-overflow hypergraph.}
\label{fig:s1}
\end{figure}

\begin{figure}[t]
    \centering
    \includegraphics[width=1.0\linewidth]{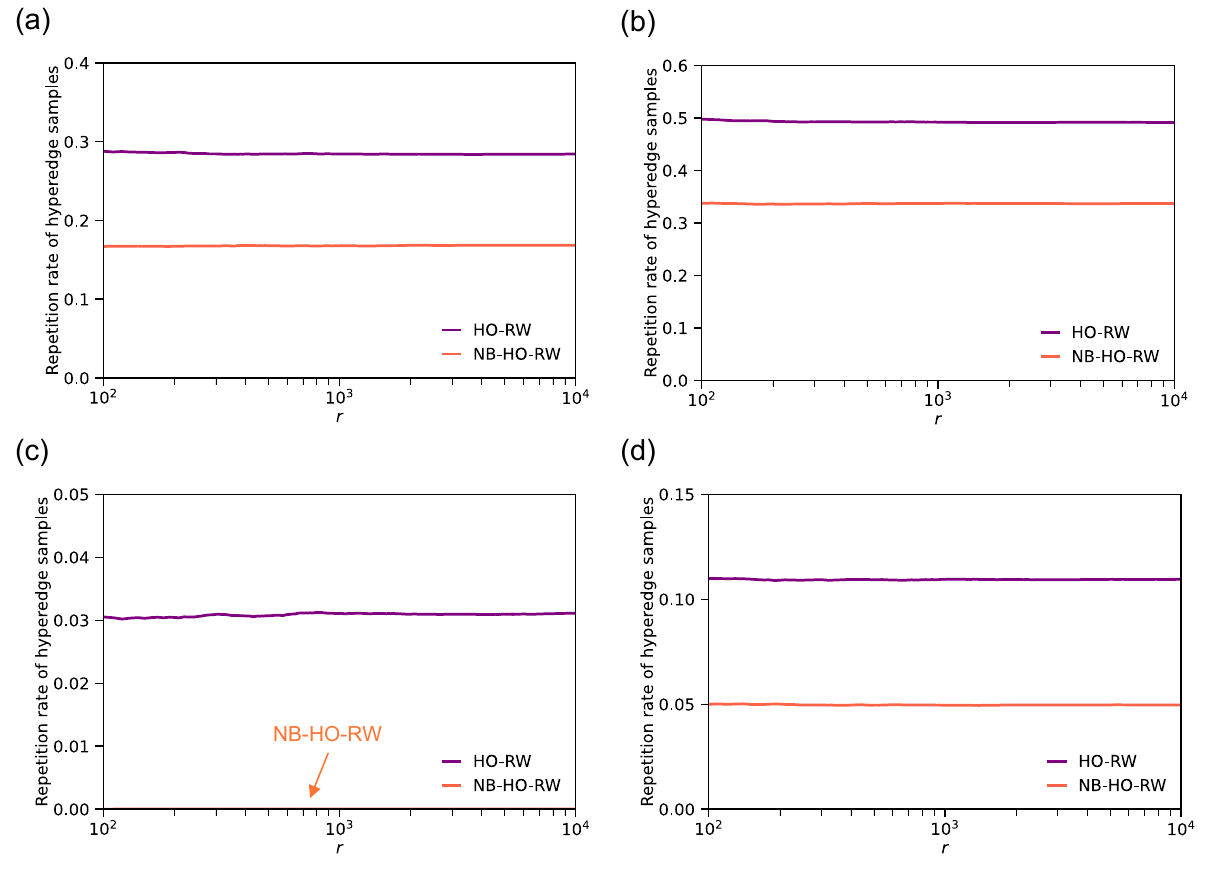}\\
    \caption{Comparison of the repetition rate of hyperedge samples. (a) MAG-geology hypergraph. (b) MAG-history hypergraph. (c) Amazon hypergraph. (d) stack-overflow hypergraph.}
\label{fig:s2}
\end{figure}

\begin{figure}[t]
    \centering
    \includegraphics[width=1.0\linewidth]{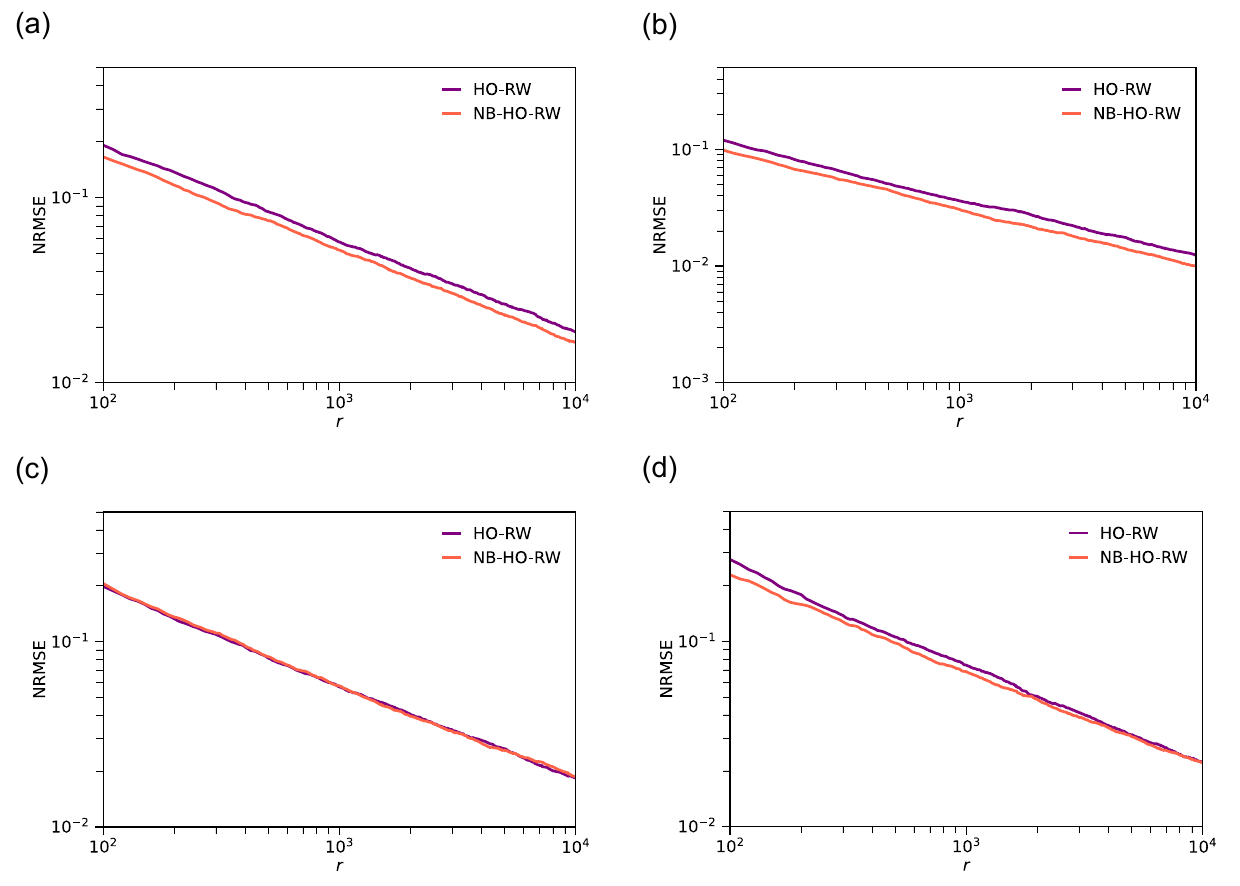}\\
    \caption{NRMSE of the estimator of the average node degree as a function of $r$. (a) MAG-geology hypergraph. (b) MAG-history hypergraph. (c) Amazon hypergraph. (d) stack-overflow hypergraph.}
\label{fig:s3}
\end{figure}

\begin{figure}[t]
    \centering
    \includegraphics[width=1.0\linewidth]{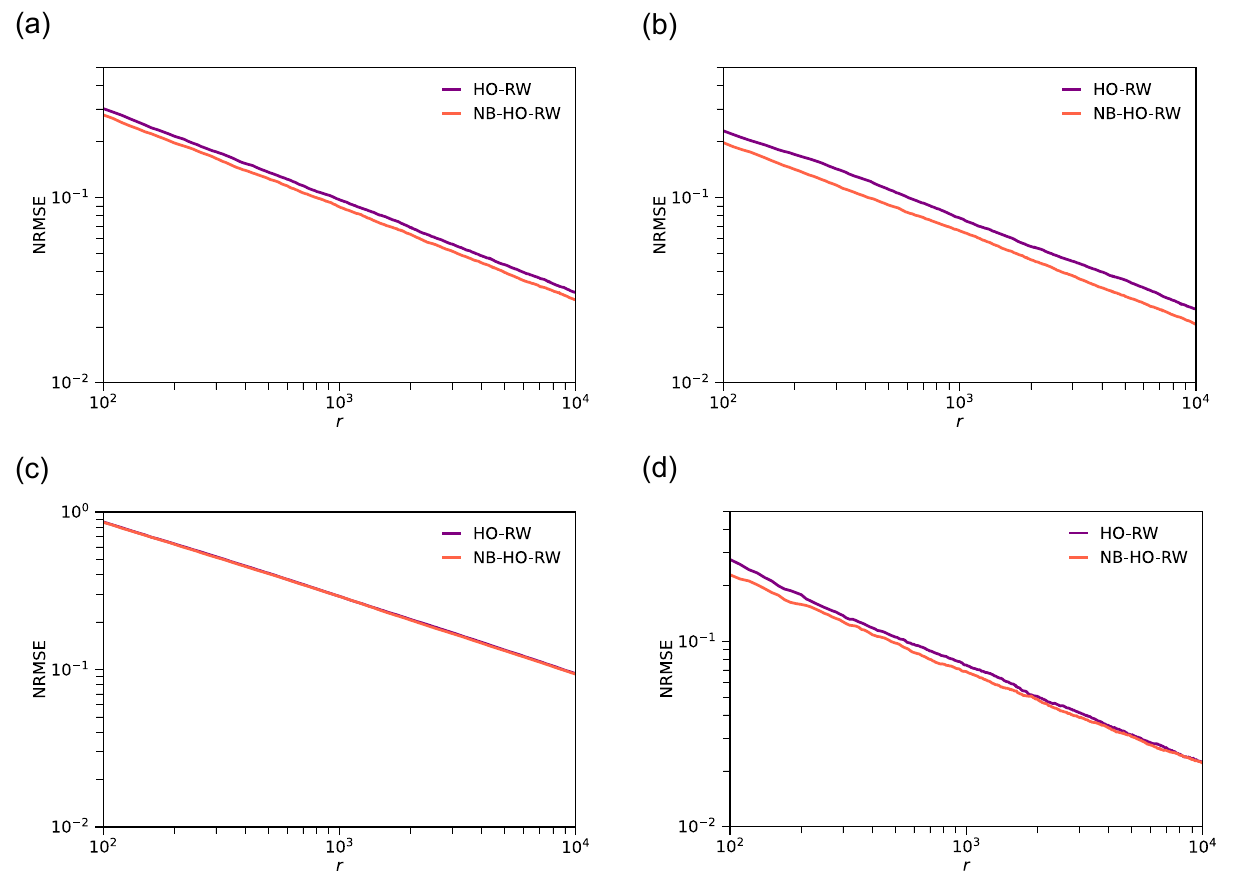}\\
    \caption{NRMSE of the estimator of the node degree distribution as a function of $r$. (a) MAG-geology hypergraph. (b) MAG-history hypergraph. (c) Amazon hypergraph. (d) stack-overflow hypergraph.}
\label{fig:s4}
\end{figure}

\begin{figure}[t]
    \centering
    \includegraphics[width=1.0\linewidth]{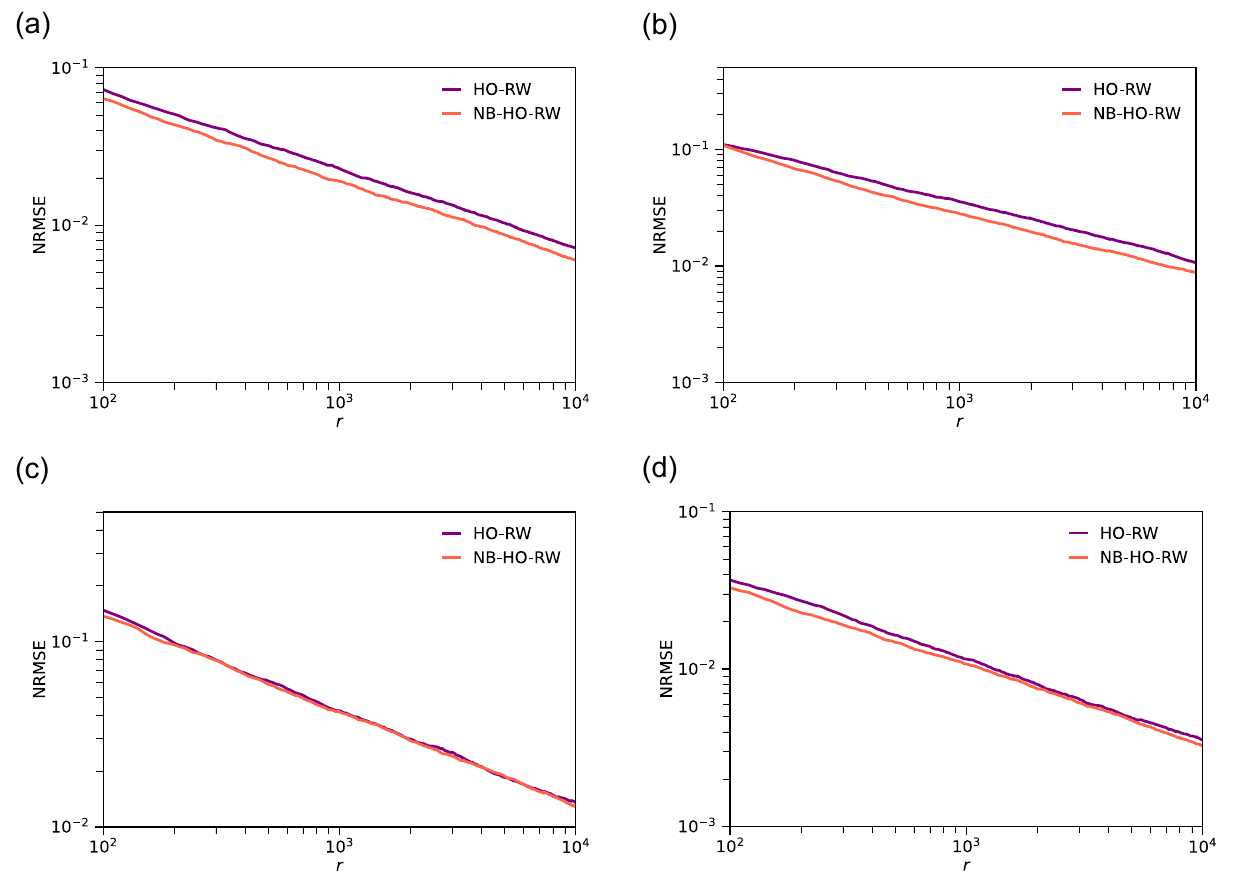}\\
    \caption{NRMSE of the estimator of the average hyperedge size as a function of $r$. (a) MAG-geology hypergraph. (b) MAG-history hypergraph. (c) Amazon hypergraph. (d) stack-overflow hypergraph.}
\label{fig:s5}
\end{figure}

\begin{figure}[t]
    \centering
    \includegraphics[width=1.0\linewidth]{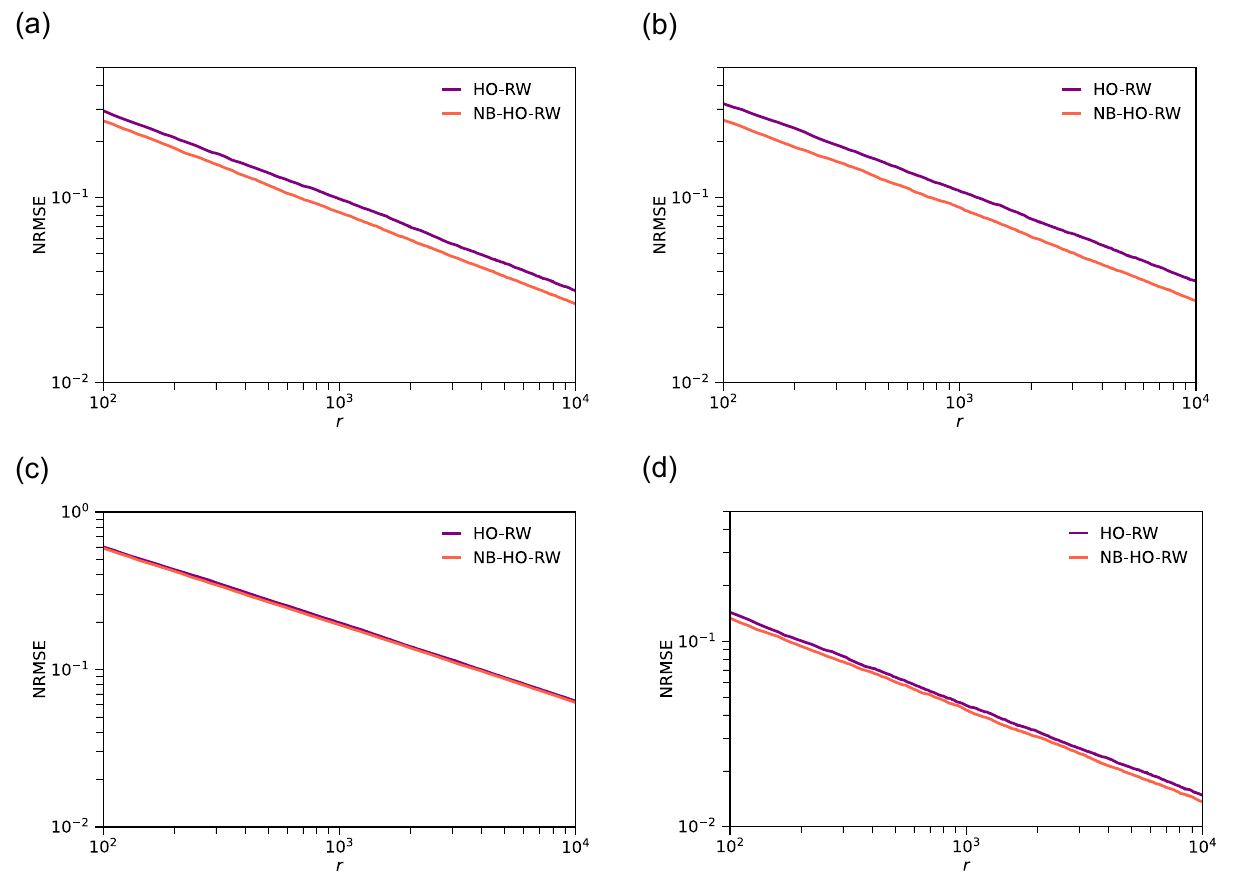}\\
    \caption{NRMSE of the estimator of the hyperedge size distribution as a function of $r$. (a) MAG-geology hypergraph. (b) MAG-history hypergraph. (c) Amazon hypergraph. (d) stack-overflow hypergraph.}
\label{fig:s6}
\end{figure}

\begin{figure}[t]
    \centering
    \includegraphics[width=0.85\linewidth]{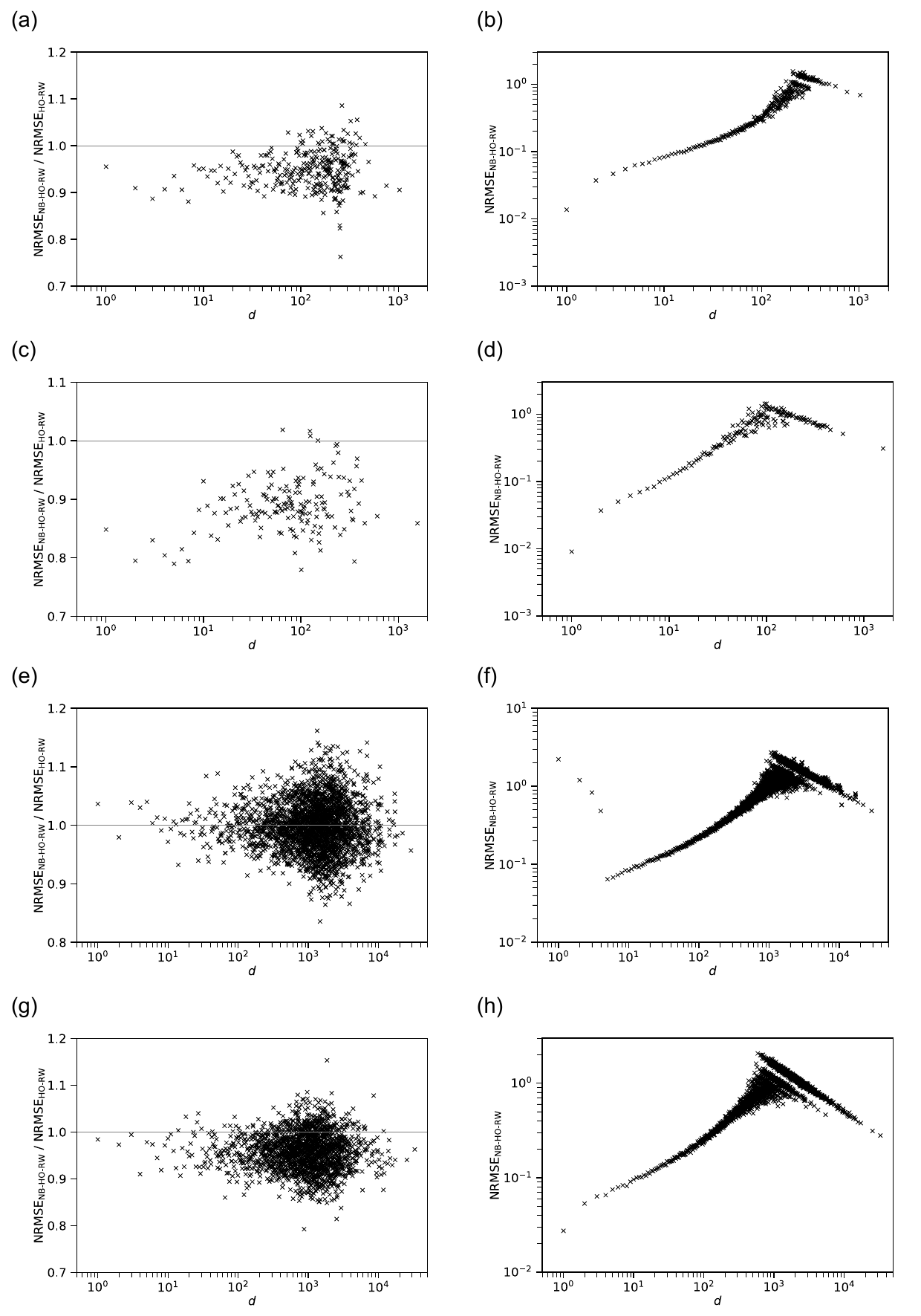}\\
    \caption{NRMSE of the estimator of the probability that a node has degree $d$ for $r = 10^4$. (a) and (b) MAG-geology hypergraph. (c) and (d) MAG-history hypergraph. (e) and (f) Amazon hypergraph. (g) and (h) stack-overflow hypergraph.}
\label{fig:s7}
\end{figure}

\begin{figure}[t]
    \centering
    \includegraphics[width=0.85\linewidth]{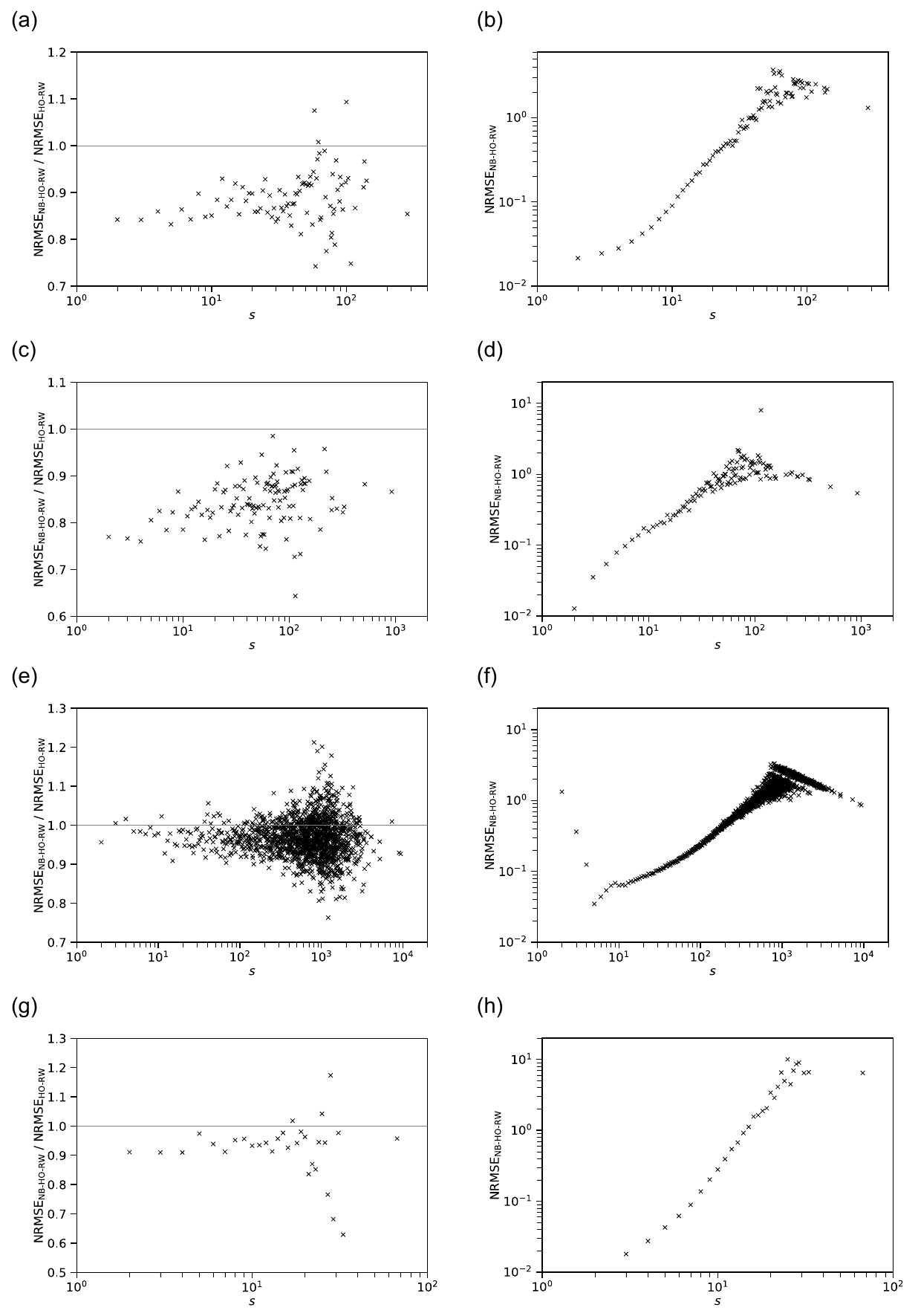}\\
    \caption{NRMSE of the estimator of the probability that a hyperedge has size $s$ for $r = 10^4$. (a) and (b) MAG-geology hypergraph. (c) and (d) MAG-history hypergraph. (e) and (f) Amazon hypergraph. (g) and (h) stack-overflow hypergraph.}
\label{fig:s8}
\end{figure}

We show the simulation results for other hypergraphs.
Figure \ref{fig:s1} shows the number of queries generated to hyperedges as a function of $r$.
Figure \ref{fig:s2} shows the repetition rate of hyperedge samples as a function of $r$.
Figure \ref{fig:s3} shows the NRMSE of the estimator of the average node degree as a function of $r$.
Figure \ref{fig:s4} shows the NRMSE of the estimator of the node degree distribution as a function of $r$.
Figure \ref{fig:s5} shows the NRMSE of the estimator of the average hyperedge size as a function of $r$.
Figure \ref{fig:s6} shows the NRMSE of the estimator of the hyperedge size distribution as a function of $r$.
Figure \ref{fig:s7} shows the NRMSE of the estimator of the probability that a node has degree $d$ as a function of $d$ for $r = 10^4$. 
Figure \ref{fig:s8} shows the NRMSE of the estimator of the probability that a hyperedge has size $s$ as a function of $s$ for $r = 10^4$. 

\renewcommand{\refname}{Supplementary References}

\end{document}